\documentstyle[epsfig]{mn}
\input{epsf}

\renewcommand{\d}{{\rm d}}
\newcommand{\beq}{\begin{equation}}
\newcommand{\eeq}{\end{equation}}
\newcommand{\beqa}{\begin{eqnarray}} 
\newcommand{\eeqa}{\end{eqnarray}}
\newcommand{\bea}{\begin{array}} 
\newcommand{\ea}{\end{array}} 
\newcommand{\lag}{\langle}
\newcommand{\rag}{\rangle}
\newcommand{\Om}{\Omega_{\rm m}}
\newcommand{\Ol}{\Omega_{\Lambda}}
\newcommand{\cP}{{\cal P}}
\newcommand{\inta}{\int_{-i\infty}^{+i\infty}}
\newcommand{\rhob}{\overline{\rho}}
\newcommand{\xib}{\overline{\xi}}
\newcommand{\rhoR}{\rho_R}
\newcommand{\bx}{{\bf x}}

\newcommand{\deltaR}{\delta_R}
\newcommand{\R}{\Re}

\newcommand{\varphih}{\hat{\varphi}}
\newcommand{\zetah}{\hat{\zeta}}
\newcommand{\tauh}{\hat{\tau}}
\newcommand{\cF}{{\cal F}}
\newcommand{\SQL}{S_3^{\rm QL}}
\newcommand{\SHEPT}{S_3^{\rm HEPT}}
\newcommand{\neff}{n_{\rm eff}}
\newcommand{\nNL}{n_{\rm NL}}
\newcommand{\Pln}{\cP_{\rm ln}}
\newcommand{\Nb}{\overline{N}}
\newcommand{\nb}{\overline{n}}

\title[A new analytical model for the evolution of the cosmological density 
distribution function]
{Evolution of the Cosmological Density Distribution Function: 
A New Analytical Model}
\author[P. Valageas \& D. Munshi]
{Patrick Valageas$^{1}$, Dipak Munshi$^{2,3}$ \\
$^{1}$Service de Physique Th\'eorique, 
CEA Saclay, 91191 Gif-sur-Yvette, France \\
$^{2}$Institute of Astronomy, Madingley Road,
Cambridge, CB3 OHA, United Kingdom\\
$^{3}$Astrophysics Group, Cavendish Laboratory, Madingley Road, 
Cambridge CB3 OHE, United Kingdom\\
}

\begin{document}
\maketitle

\begin{abstract}
The one-point probability distribution function (pdf) of the large-scale 
density field is an important tool to follow the evolution of cosmological
structures. In this paper we present a new model for this pdf for all regimes
and all densities, that is from linear to highly non-linear scales and from
rare voids up to rare high densities. This is probably the simplest model
one can build which is consistent with normalization constraints and known 
rigorous results (the quasi-linear regime and the rare void limit). It is fully
parameterized by the non-linear variance and skewness. We obtain a good
agreement with N-body data from realistic cosmological simulations of
the VIRGO consortium and we find that it works significantly better than 
previous models such as the lognormal model or the Extended Perturbation 
Theory (EPT). We explain this success as a result of the tight
constraints onto the pdf provided by these consistency conditions. We also
point out that while the Lagrangian dynamics of typical fluctuations is quite
complex the statistical outcome seems rather simple. This simple model should 
be useful for studies which require a realistic and convenient description
of this pdf.
\end{abstract}


\section{Introduction}

In usual cosmological scenarios the large-scale structures observed in the 
present universe have formed by gravitational instability from small initial 
density fluctuations. Moreover, in a CDM-like model the system is governed
at large scales by the collisionless gravitational dynamics of the dark matter
which builds a wide variety of structures, from rare expanding voids to
filaments and almost spherical massive clusters. Besides, in such models
where the amplitude of density fluctuations grows at smaller scales one
observes the build-up of a hierarchical clustering process. The scale which
marks the transition to non-linearity increases with time so that increasingly
large and massive structures turn non-linear and collapse as time goes on and
smaller objects which formed earlier become embedded within larger entities.
Then, they may follow a complex dynamics as mergings and disruptions take 
place. This builds a complex network (or cosmic web) with a wealth of 
structures and substructures which is quite difficult to model in details.
Then, at small scales where the baryonic density is sufficiently large 
(e.g., within collapsed halos) radiative processes like cooling come into play
and further enhance the gas density through the subsequent baryonic collapse
they imply. The baryonic matter may become a dominant component and eventually
form stars and galaxies. Therefore, the large scale distribution of matter
is closely related to the observed distribution of astrophysical objects
(galaxies, clusters, Lyman-$\alpha$ clouds, etc.) although on very small scales
the relationship can become very intricate. Thus, an important goal of 
observational cosmology is to understand the formation of these large-scale 
structures.

A first step is to investigate the distribution of matter on large scales
which is governed by gravity (note that the density field is also directly 
probed by weak gravitational lensing surveys as opposed to galaxy surveys
which probe the stellar content of the universe). One may then study the 
mass function of specific
objects, like just-virialized halos, following Press \& Schechter (1974).
Another tool is provided by the many-body density correlation functions
which probe in more details the structure of the density field (e.g., Peebles
1980). In this article we shall focus on a simpler statistics: the one-point 
probability distribution function (pdf) of the density $\cP(\rhoR)$. Although
it discards the angular behaviour of the many-body correlations it contains
their amplitude at a given scale $R$ and it allows one to follow the evolution
of gravitational clustering with time. Indeed, by going from low to high 
densities one successively probes rare voids, filaments, typical halos and
rare massive halos (identified with clusters at $z=0$). We shall assume here
that the initial conditions are Gaussian, as expected from simple inflationary
scenarios. Then, at large scales or at high redshifts the pdf goes to a simple
Gaussian. Next, as non-linearities develop and gravitational clustering builds
up deviations from Gaussianity appear and become increasingly important.

The first deviations from Gaussianity can be derived from analytical methods
in the so-called quasi-linear limit, from perturbative expansions (Bernardeau
1992, 1994a) or steepest-descent methods (Valageas 2002a). In a similar 
fashion, the statistics of extreme underdensities (rare voids) can also be
obtained for any regime (Valageas 2002a, see also Bernardeau 1994b). However,
it has proved difficult to model the evolution of the full pdf $\cP(\rhoR)$
into the linear regime. It was proposed to use as a simple approximation
a lognormal pdf (Coles \& Jones 1991) which is always well-behaved as the 
density is positive (while the Gaussian pdf extends to increasingly negative 
densities in the non-linear regime). However, this model is inconsistent with
the known results and it appears to show significant discrepancies with
detailed numerical simulations (Bernardeau \& Kofman 1995). 

A more theoretical
approach has been to investigate a Lagrangian method where one tries to follow
the evolution of individual density fluctuations. This can be expressed in 
terms of a mapping between the linear density contrast and the
actual non-linear density contrast. As seen in Valageas (1998) and 
Protogerros \& Scherrer (1997), using the mapping given by the spherical
dynamics yields the exact quasi-linear limit and provides good results up
to a rms linear density fluctuation $\sigma \la 1$. This can be understood
in simple terms from the steepest-descent method developed in 
Valageas (2002a). A detailed study of the spherical model, compared with
perturbation theory, various local approximations and numerical simulations,
is presented in Fosalba \& Gaztanaga (1998). A different approach based on the
excursion set formalism which also includes the spherical model is described
in Sheth (1998). However, the highly non-linear regime where mergings and 
tidal disruptions play a key role seems beyond the reach of such approaches 
(at least so far). Other approaches to tackle non-linear gravity include
semi-analytical techniques which approximate the Vlasov-Poisson dynamics 
by simpler systems of equations (for a comparison of such models in the
weakly non-linear regime see Munshi et al. 1994).

Since following the dynamics of individual density fluctuations appears
problematic in the non-linear regime (because of mergers, etc.) other 
approaches have been advocated which directly consider the statistical 
properties of the system. For instance, assuming specific scaling-laws
for the many-body correlation functions inspired from the stable-clustering
Ansatz (see Peebles 1980), Balian \& Schaeffer (1989) were able to derive
the many properties they imply for the density field. Note that the formalism
developed in that work can be used in more general cases (like the use of
cumulant generating functions which we also take advantage of in this paper).
Then, Scoccimarro \& Frieman (1999) devised a model (HEPT) to obtain the
amplitude of these many-body correlations in the highly non-linear regime
(assuming again that they obey the same scaling laws). They proposed to
extrapolate to the non-linear regime the tree-level perturbation results
through a specific limit and obtained a good agreement with numerical
simulations.

A somewhat more empirical attempt to follow the evolution of gravitational 
clustering was proposed in Colombi et al. (1997). Working also with the pdf 
$\cP(\rhoR)$ itself (or more precisely its cumulant generating function 
$\varphi(y)$) they modeled the non-linear pdf through a simple empirical 
modification of the quasi-linear prediction, which they dubbed Extended 
Perturbation Theory (EPT). More precisely, they suggested to use the functional
shape derived in the quasi-linear limit but to treat the local slope $n$ of
the power-spectrum as a free parameter in order to extend the model into the
non-linear regime. They obtained in this manner a reasonable agreement with 
numerical simulations. 
In this article we follow the same approach as we directly work
with the generating function $\varphi(y)$ but in addition to the quasi-linear
limit we also require that it satisfies the rare-void limit. Then, we build
the simplest possible model which obeys all these constraints. As EPT it
is fully parameterized by the non-linear variance and the skewness. Using 
standard models for these two quantities we show through a comparison with 
numerical simulations that our approach provides a good prediction for the 
pdf in all regimes. In particular, it works significantly better than both 
the lognormal approximation and EPT.

This paper is organized as follows. In section~\ref{Models} we present our 
model and we recall both the EPT and lognormal models. Then, we compare our 
results to N-body simulations in section~\ref{Simulations} and we conclude 
in section~\ref{Conclusion}. We also explain the numerical implementation
of our model in appendix~\ref{Numerical implementation} and we discuss its 
robustness in appendix~\ref{Robustness of the model}.

\section{Models for the density distribution function}
\label{Models}

\subsection{General framework}
\label{General framework}

In this article we wish to build a simple model for the one-point probability 
distribution function (pdf) of the density, which describes the formation of
large scale structures in the universe through gravitational instability.
We first define the overdensity at scale $R$ as the mean overdensity over
a spherical cell of radius $R$ and volume $V$ (i.e. we use a spherical top-hat 
filter):
\beq
\rhoR= \int_V \frac{\d\bx}{V} \; \frac{\rho(\bx)}{\rhob} = 1 + \deltaR ,
\label{rhoR}
\eeq
where $\rhob$ is the mean density of the universe, $\deltaR$ is the
density contrast at the same scale $R$ and $\bx$ is the comoving coordinate.
The first few cumulants of the overdensity obey:
\beq
\lag 1\rag=1 , \;\; \lag\rhoR\rag=1 , \;\; \lag\rhoR^2\rag_c= \xib \;\;\; 
\mbox{and} \;\;\; \frac{\lag\rhoR^3\rag_c}{\xib^2} = S_3 .
\label{XiS3}
\eeq
Here we noted $\lag..\rag$ the average over the initial Gaussian conditions
(or over space by ergodicity) while the subscript $c$ refers to ``cumulants''
(or connected parts) as opposed to simple moments. We also introduced in
eq.(\ref{XiS3}) the variance $\xib$ and the skewness $S_3$ while the first
constraint in (\ref{XiS3}) simply means that the pdf $\cP(\rhoR)$ is 
normalized to unity. A Gaussian
density field is fully defined by its variance and all higher-order cumulants
vanish (in particular $S_3=0$). As is well-known, although we assume the 
initial conditions to be Gaussian the non-linear gravitational dynamics
will gradually build non-Gaussianities which we need to take into account.
Therefore, we introduce the cumulant generating function $\varphi(y)$ defined 
by:
\beq
\varphi(y) = \sum_{p=0}^{\infty} \frac{(-1)^{p-1}}{p!} \; S_p \; y^p ,
\label{phi}
\eeq
with:
\beq
S_0=0 , \;\; S_1= S_2=1 \;\;\; \mbox{and for } p \geq 3 : \; 
S_p = \frac{\lag\rhoR^p\rag_c}{\xib^{\;p-1}} .
\label{Sp}
\eeq
Thus, all cumulants can be obtained from $\varphi(y)$ (assuming the series
in eq.(\ref{phi}) converges). Moreover, this generating function is simply
the logarithm of the Laplace transform of the pdf $\cP(\rhoR)$:
\beq
e^{-\varphi(y)/\xib} = \int_0^{\infty} \d\rhoR \; 
e^{-\rhoR y/\xib} \; \cP(\rhoR) ,
\label{phiP}
\eeq
which can be inverted as:
\beq
\cP(\rhoR) = \inta \frac{\d y}{2\pi i \xib} \; 
e^{[\rhoR y - \varphi(y)] /\xib} .
\label{Pphi}
\eeq
One advantage of working with the generating function $\varphi(y)$ rather
than with the pdf $\cP(\rhoR)$ itself is that it provides a very simple and
flexible tool to parameterize the non-linear evolution of gravitational
clustering. For instance, the four normalization constraints (\ref{XiS3})
are automatically satisfied if we ensure that the Taylor expansion at $y=0$
of $\varphi(y)$ is: $\varphi(y)=y-y^2/2+S_3y^3/6-..$. It is easier to
implement such local conditions than the integral constraints (\ref{XiS3}) 
onto the pdf $\cP(\rhoR)$ itself. However, one must also check that the 
pdf obtained from eq.(\ref{Pphi}) is positive. More precisely, we have:
\beq
\cP(\rhoR) \geq 0 \; \mbox{for} \; \rhoR \geq 0 \;\;\; \mbox{and} \;\;\;
\cP(\rhoR) = 0 \; \mbox{for} \; \rhoR < 0 .
\label{positive}
\eeq
As seen from eq.(\ref{Pphi}), the second constraint in (\ref{positive})
is satisfied as soon as $\varphi(y)$ is regular over the right half-plane
$\R(y) \geq 0$ and grows more slowly than $|y|$ for $\R(y) \rightarrow 
+\infty$.

\subsection{Quasi-linear regime}
\label{Quasi-linear regime}

The functions $\cP(\rhoR)$ and $\varphi(y)$ can be derived in a rigorous 
manner at large scales or at early times where the amplitude of the density
fluctuations goes to zero. In this quasi-linear regime, it is convenient
to introduce the rms linear density fluctuation $\sigma$:
\beq
\sigma^2 = \lag \delta_{L,R}^2 \rag ,
\label{sigma}
\eeq
where $\delta_{L,R}$ is the linear density contrast smoothed at scale $R$.
Of course, at large scales we have $\xib/\sigma^2 \rightarrow 1$. Then, the
appropriate generating function $\varphih(y)$ is now defined as:
\beq
e^{-\varphih(y)/\sigma^2} = \int_0^{\infty} \d\rhoR \; 
e^{-\rhoR y/\sigma^2} \; \cP(\rhoR) .
\label{phihP}
\eeq
Thus, it is related to $\varphi(y)$ defined in eq.(\ref{phiP}) by:
\beq
\varphi(y) = \frac{\xib}{\sigma^2} \; 
\varphih \left( \frac{\sigma^2}{\xib} \; y \right) .
\label{phiphih}
\eeq
Again, at large scales we have $\varphi(y) \rightarrow \varphih(y)$. It turns
out that the function $\varphih(y)$ has a finite limit at fixed $y$ in the
quasi-linear limit $\sigma^2 \rightarrow 0$ which can be computed 
analytically. This may be done through a perturbative expansion of the
density field (Bernardeau 1992, 1994a) or a steepest-descent method 
(Valageas 2002a). In both cases, one obtains the cumulant generating function
$\varphih(y)$ as the solution of the implicit system:
\beqa
\varphih(y) & = & y \zetah(\tauh) + \frac{\tauh^2}{2}
\label{phihzetah} \\
\tauh & = & - y \; \zetah'(\tauh)
\label{ytauh}
\eeqa
where the function $\zetah(\tauh)$ is closely related to the spherical 
collapse. Indeed, it is defined by the implicit relation:
\beq
\zetah(\tauh) = \cF\left[ -\tauh \; \frac{\sigma\left[\zetah(\tauh)^{1/3}
R\right]}{\sigma(R)} \right] ,
\label{zeta}
\eeq
where the function $\cF$ describes the mapping from the linear density
contrast $\delta_L$ to the non-linear overdensity $\rhoR$ given by the
spherical collapse:
\beq
\rhoR=\cF(\delta_L) \;\; \mbox{before shell-crossing} .
\label{rhoF}
\eeq
In the case of a critical-density universe, the function $\cF(\delta_L)$
can be written in terms of trigonometric and hyperbolic functions 
(Peebles 1980). However, it turns out that the dependence of $\cF$ on the
cosmology is quite weak and that very good results can be obtained by using
its limit for $\Om \rightarrow 0$, with $\Ol=0$, (see Bernardeau 1992):
\beq
\cF(\delta_L) = \left( 1 - \frac{2}{3} \delta_L \right)^{-3/2} .
\label{F}
\eeq
Moreover, the power-law behaviour $\cF(\delta_L) \propto 
(-\delta_L)^{-3/2}$ for $\delta_L \rightarrow -\infty$ is exact for
any values of $\Om$ and $\Ol$. The implicit equation (\ref{zeta}) can be
simplified for a power-law linear power-spectrum $P_L(k) \propto k^n$. We
shall assume hereafter that we have $-3<n<1$, which agrees with usual 
cosmological models like CDM power-spectra. In this case, we have:
\beq
\sigma^2(R) \propto R^{-(n+3)} \;\; \mbox{and also:} \;\; 
n+3= - \frac{\d\ln\sigma^2}{\d\ln R} ,
\label{nsigma}
\eeq
and eq.(\ref{zeta}) writes:
\beq
\zetah = \cF\left[ -\tauh \; \zetah^{-(n+3)/6} \right] .
\label{zetan}
\eeq
Using the expression (\ref{F}) for $\cF$ we obtain the inverse $\tauh(\zetah)$
as:
\beq
\tauh(\zetah) = - \frac{3}{2} \zetah^{(n+3)/6} + \frac{3}{2} \zetah^{(n-1)/6} .
\label{tauhzetah} 
\eeq
This equation fully defines $\varphi(y)=\varphih(y)$ in the quasi-linear
regime, through eqs.(\ref{phihzetah})-(\ref{ytauh}), whence the pdf 
$\cP(\rhoR)$ from eq.(\ref{Pphi}). We must note that in studies which focus
on the quasi-linear regime one often considers the density contrast $\deltaR$
rather than the overdensity $\rhoR$. This yields an additional factor $-1$
to $\cF$ and $\zetah$ while $\varphih(y)$ is defined with $S_1=0$.

\subsection{Rare voids}
\label{Rare voids}

In addition to the quasi-linear regime, one can also derive exact results 
for any $\sigma^2$ in the limit of extreme underdensities. This corresponds
to the limit of large positive $y$ and $\tauh$ and small $\zetah$ at fixed
$\sigma$. This is again done through a steepest-descent method which is 
appropriate to rare events (Valageas 2002b). The spherical saddle-point of 
the relevant action is actually the same as the one obtained in the 
quasi-linear regime so that the generating function $\varphih(y)$ is still 
given by the implicit system (\ref{phihzetah})-(\ref{ytauh}) while the
function $\tauh(\zetah)$ agrees with eq.(\ref{tauhzetah}):
\beq
\tauh(\zetah) \simeq \frac{3}{2} \zetah^{(n-1)/6} \;\;\; \mbox{for} \;\;\;
\tauh \gg \sigma \;\;\; \mbox{and any} \; \sigma .
\label{tauhvoid}
\eeq
Note that in Valageas (2002b) the factor $3/2$ was replaced by $27/20$ since
we considered a critical-density universe (this again shows that the dependence
of $\tauh(\zetah)$ on cosmology is rather weak). As explained in Valageas 
(2002b), the asymptotic behaviour (\ref{tauhvoid}) holds for rare 
underdensities whatever the value of $\sigma$ (hence it remains valid in the
highly non-linear regime). The reason why the generating function 
$\varphih(y)$ can still be written in terms of the spherical dynamics 
$\cF(\delta_{L,R})$ is that rare voids become increasingly spherical as they
expand and they are not affected by shell-crossing yet. Of course, if one
follows the evolution of a particular void after some time its expansion will
be stopped or slowed down when it encounters more extreme voids (while 
filaments will form at their boundaries). This simply means that such a void
is no longer part of the rare events described by eq.(\ref{tauhvoid}) which
applies to increasingly rare (but also more extended) voids as time goes on.
As noticed in Valageas (2002b), for steep linear power-spectra $n>-1$ the
radial profile of the spherical saddle-point which led to eq.(\ref{tauhvoid})
shows some shell-crossing at radius $R$. Although this behaviour is likely to
modify the numerical factor $3/2$ in eq.(\ref{tauhvoid}) we can expect the 
exponent to remain correct for $n>-1$ (on the other hand, note that for 
CDM-like power-spectra we have $n \la -1$ at the scales of interest).

\subsection{A simple model}
\label{A simple model}

Gathering the results recalled in the previous sections we now build a simple
model to describe the evolution of the pdf $\cP(\rhoR)$. Of course, we
wish to recover both the quasi-linear regime presented in 
section~\ref{Quasi-linear regime} and the rare underdensities regime recalled
in section~\ref{Rare voids}. From eq.(\ref{phiphih}), we can see that if
the generating function $\varphih(y)$ is given by an implicit system of the
form (\ref{phihzetah})-(\ref{ytauh}), then the generating function 
$\varphi(y)$ is defined by the same system where $\zetah(\tauh)$ is replaced
by $\zeta(\tau)$ with:
\beq
\zeta(\tau) = \zetah \left( \sqrt{\frac{\sigma^2}{\xib}} \; \tau \right) .
\label{zetazetah}
\eeq
Since in the quasi-linear regime the inverse $\tauh(\zetah)$ has a simpler
expression than $\zetah(\tauh)$, see eq.(\ref{tauhzetah}), we work with
$\tau(\zeta)$ and we parameterize the generating function $\varphi(y)$ by:
\beqa
\varphi(y) & = & y \zeta + \frac{\tau^2}{2}
\label{phizeta} \\
y & = & - \tau \frac{\d\tau}{\d\zeta}
\label{ytau}
\eeqa
In fact, we see from eqs.(\ref{phizeta})-(\ref{ytau}) that $-\tau^2(\zeta)/2$
is merely the Legendre transform of $\varphi(y)$. Next, the quasi-linear
limit implies from eq.(\ref{tauhzetah}) and eq.(\ref{zetazetah}):
\beq
\sigma \rightarrow 0 : \;\; \tau(\zeta) \rightarrow 
- \frac{3}{2} \zeta^{(n+3)/6} + \frac{3}{2} \zeta^{(n-1)/6} ,
\label{tauzetaQL} 
\eeq
while the underdense limit implies from eq.(\ref{tauhvoid}) and 
eq.(\ref{zetazetah}):
\beq
\zeta \rightarrow 0 : \;\; \tau(\zeta) \rightarrow \frac{3}{2} 
\sqrt{\frac{\xib}{\sigma^2}} \; \zeta^{(n-1)/6} .
\label{tauvoid}
\eeq
In addition to these asymptotic behaviours, we also have the normalization
constraints (\ref{XiS3}) or (\ref{Sp}). From eqs.(\ref{phizeta})-(\ref{ytau})
it is easy to see that the conditions (\ref{XiS3}) are equivalent to the
following constraints for $\tau(\zeta)$ (note that $y=0$ corresponds to
$\tau=0$ and $\zeta=1$):
\beq
\tau(1) = 0, \;\; \tau'(1)= -1 \;\; \mbox{and} \;\; \tau''(1)=\frac{S_3}{3} .
\label{taunorm}
\eeq
Indeed, note that the first normalization condition $\lag 1\rag=1$ in 
(\ref{XiS3}), or $\varphi(0)=0$, is automatically satisfied by the implicit
system (\ref{phizeta})-(\ref{ytau}) for any function $\tau(\zeta)$, provided
the transform $\varphi(y) \leftrightarrow \tau(\zeta)$ it defines is regular 
around $(y=0,\varphi=0) \leftrightarrow (\zeta=1,\tau=0)$. In order to
obey the constraints (\ref{tauzetaQL})-(\ref{taunorm}) we consider the simple
model:
\beq
\tau(\zeta) = a + b \; \zeta^2 + c \; \zeta^{(n+3)/6} + \frac{3}{2} 
\sqrt{\frac{\xib}{\sigma^2}} \; \zeta^{(n-1)/6} ,
\label{tauzeta}
\eeq
where the coefficients $a,b,c$ are set by the constraints (\ref{taunorm}).
This is consistent with the underdense limit (\ref{tauvoid}) and we recover 
the quasi-linear limit (\ref{tauzetaQL}) provided we use for the skewness
$S_3$ the appropriate perturbative prediction $\SQL$:
\beq
\sigma \rightarrow 0 : \;\; S_3 \rightarrow \SQL = 5 - (n+3) ,
\label{S3QL}
\eeq
which implies with (\ref{taunorm}):
\beq
\sigma \rightarrow 0 : \;\; a \rightarrow 0 , \;\; b \rightarrow 0 , 
\;\; c \rightarrow -\frac{3}{2}.
\eeq
Note that one often chooses the value associated with a critical-density
universe $\SQL=34/7-(n+3)$ rather than the result (\ref{S3QL}) obtained for
a ``zero-density'' open universe. We prefer to use (\ref{S3QL}) which is
consistent with (\ref{tauhzetah}) and quite sufficient for our purposes.
In other words, we neglect the small dependence of the skewness on 
cosmological parameters. Using eq.(\ref{taunorm}) we obtain $a, b$ and $c$ 
from:
\beq
c= - \frac{12 S_3+36-(1-n)(13-n)\frac{3}{2}\sqrt{\frac{\xib}{\sigma^2}}}
{(n+3)(9-n)} ,
\label{c}
\eeq
\beq
b= \frac{1-n}{12}\frac{3}{2}\sqrt{\frac{\xib}{\sigma^2}} - \frac{1}{2}
- \frac{n+3}{12} c ,
\label{b}
\eeq
and:
\beq
a=-b-c-\frac{3}{2}\sqrt{\frac{\xib}{\sigma^2}} .
\label{a}
\eeq

\subsection{Variance and skewness}
\label{Variance and skewness}

Finally, in order to fully determine the pdf $\cP(\rhoR)$ we need the
variance $\xib$ and the skewness $S_3$. For the variance we simply use
the fit provided by Peacock \& Dodds (1996) for the evolution of the
non-linear power-spectrum $P(k)$. However, one could as well use any other
model which fits the data, like those presented in Smith et al. (2003).  
The fitting functions presented in Peacock \& Dodds (1996) for $P(k)$ depend 
on the slope $n(k_L/2)$ of the linear power-spectrum $P_L(k_L)$ at the 
``Lagrangian'' wavenumber $k_L/2$ (see their article for details). For the 
numerical results presented below in section~\ref{Simulations} for a 
$\Lambda$CDM model, we define
this local index $n$ from the logarithmic derivative of the linear 
power-spectrum at wavenumber $k_L$ rather than $k_L/2$. This provides a better
fit to the simulations we study here and it shows the correct limit at large 
scales (i.e. $k_L \rightarrow k$).

Next, we need to specify the skewness $S_3$. We choose a simple interpolation
between the quasi-linear prediction $\SQL$ and the highly non-linear
prediction $\SHEPT$ of the Hyper Extended Perturbation Theory (HEPT) 
presented in Scoccimarro \& Frieman (1999):
\beq
S_3 = \frac{\SQL+\xib^{1.5}\SHEPT}{1+\xib^{1.5}} \;\;\; \mbox{with} \;\;\;
\SHEPT = 3 \frac{4-2^n}{1+2^{n+1}} ,
\label{S3}
\eeq
while $\SQL$ was defined in eq.(\ref{S3QL}). The skewness again depends on
the local slope of the linear power-spectrum or the linear variance. As in
Scoccimarro \& Frieman (1999) or Colombi et al. (1997) we define the index
$n$ to be used in eq.(\ref{S3}) as the logarithmic slope of the linear
variance $\sigma^2$, as in eq.(\ref{nsigma}), at the Eulerian scale $R$.
Note that this procedure is different from the one used to obtain the
non-linear power-spectrum (which follows Peacock \& Dodds 1996) but in
agreement with previous works we found that it yields better results for the
skewness. Next, we also choose for the index $n$ which appears explicitly in 
eq.(\ref{tauzeta}) the value we use for the skewness, obtained from 
eq.(\ref{nsigma}).

It may seem a bit surprising
to use a different prescription for $n$ for the variance and the skewness
but this should be seen as a result of the uncertainty on the variance $\xib$
(i.e. the non-linear power-spectrum). For instance, the fitting formulae
for $P(k)$ given by Smith et al. (2003), which are based on a halo model, 
define $n$ from the linear variance as in eq.(\ref{nsigma}) but at the 
transition scale $R_0$ where $\sigma=1$. Of course, all these prescriptions
are identical in the case of an exact power-law linear power-spectrum. From 
the point of view of this article, these points are not part of the
model we propose for the evolution of the pdf $\cP(\rhoR)$. Our model 
is fully defined by eq.(\ref{tauzeta}) and it provides
$\cP(\rhoR)$ once we are given the variance $\xib$, the skewness $S_3$ and
the local slope $n$. Thus, the user could choose any interpolation formula for
$\xib$ or $S_3$ as long as it agrees reasonably well with the data. 

For simplicity we chose to interpolate between the quasi-linear and HEPT 
predictions for $S_3$ using the simple variable $\xib^{1.5}$. From 
eq.(\ref{tauzeta}) one might have thought using the ratio $\xib/\sigma^2$.
However, this would not be convenient because the behaviour of this ratio
depends on the slope $n$ of the power-spectrum. Thus, if the stable-clustering 
Ansatz is valid this ratio would go to zero at small scales if $n>-2$ while
it would go to infinity if $n<-2$. Even though the stable-clustering Ansatz
may not be exact such a large range of behaviours can be seen in numerical
simulations (e.g., Fig.3 in Valageas et al. 2000). By contrast, $\xib>1$ 
always marks the transition to the non-linear regime. Of course, other choices
than (\ref{S3}) are possible. For instance, one could integrate the 
interpolation formula given by Scoccimarro \& Couchman (2001) for the 
bispectrum. Their interpolations are similar to eq.(\ref{S3}) except that
they use $\sigma^2$ rather than $\xib$. We tried both variables for our present
purposes and we found that $\xib$ with the power $1.5$ worked best as compared
with the numerical simulations. However, such interpolations are not fully
satisfactory as seen below in Fig.\ref{moment_skew.eps}. The skewness still
shows a rather large uncertainty in the transition regime, both from 
analytical approaches or simulation results (note that EPT was fitted to
numerical simulations).

\subsection{Other models for the pdf $\cP(\rhoR)$}
\label{Other models}

In addition to numerical simulations, we shall also compare our model 
(\ref{tauzeta}) with two other prescriptions which have been put forward in 
previous works.

\subsubsection{Extended Perturbation Theory}
\label{Extended Perturbation Theory}

In Colombi et al. (1997) it was proposed to extend the quasi-linear prediction
(\ref{tauhzetah}) into the non-linear regime by leaving $n$ as a free
parameter which deviates from the local slope of the linear variance at small
scales. Therefore, this model, dubbed Extended Perturbation Theory (EPT),
amounts to write:
\beq
\mbox{EPT} : \;\;  \tau(\zeta) = - \frac{3}{2} \zeta^{(\neff+3)/6} + 
\frac{3}{2} \zeta^{(\neff-1)/6} ,
\label{tauzetaEPT}
\eeq
where $\neff(n,\sigma)$ can be obtained from $S_3$. Colombi et al. (1997) also
gave the following fit for $\neff$:
\beq
\neff= n + \frac{x^{\alpha}(\nNL-n)}{x^{\alpha}+x^{-\alpha}} \;\;\; \mbox{with}
\;\;\; x=\exp[\log_{10}(\sigma^2/\sigma_0^2)] ,
\label{neff}
\eeq
and:
\beq
\nNL(n)= \frac{3(n-1)}{3+n} , \;\; \alpha= \frac{8-3 n}{10} , \;\; 
\log_{10}(\sigma_0^2) = \frac{2-n}{10} .
\label{nNL}
\eeq
Note that a significant difference between EPT and our model (\ref{tauzeta})
is that in the highly non-linear regime if the stable-clustering Ansatz is 
valid (whence $S_3$ is constant with time at fixed physical scale) the function
$\tau(\zeta)$ given by EPT is fixed while it keeps evolving in our model 
because of the explicit factor $\sqrt{\xib/\sigma^2}$ which translates the
continuing expansion and merging of rare voids. In particular, since the 
low-$\zeta$ limit of eq.(\ref{tauzetaEPT}) is different from the exact result
(\ref{tauhvoid}) when $\sigma \gg 1$, the EPT prediction should fail for the
low-density cutoff of the pdf $\cP(\rhoR)$ in the non-linear regime, see
section~\ref{Density pdf} below.

\subsubsection{Lognormal model}
\label{Lognormal model}

A second popular model used to describe the evolution of the pdf $\cP(\rhoR)$
is the lognormal approximation (e.g., Kayo et al. 2001). This provides a 
simple expression for the pdf $\cP(\rhoR)$ itself which reads:
\beq
\Pln(\rhoR) =
\frac{1}{\rhoR\sqrt{2\pi\ln(1+\xib)}} 
\exp \left( - \frac{\ln^2[\rhoR\sqrt{1+\xib}]}{2\ln(1+\xib)} \right) .
\label{Pln}
\eeq
This also gives for the skewness $S_3$:
\beq
S_3 = 3 + \xib .
\label{S3ln}
\eeq
Of course, in the limit $\sigma\rightarrow 0$ the lognormal pdf $\Pln(\rhoR)$
goes to the usual Gaussian. However, the coefficients $S_p$ it defines (i.e.
its cumulants) do not match the exact quasi-linear result recalled in 
section~\ref{Quasi-linear regime}. Moreover, it does not agree with the low
density limit presented in section~\ref{Rare voids}. Therefore, we can expect
a significant discrepancy with numerical simulations at the low-density
cutoff in the non-linear regime, see also section~\ref{Density pdf} below.

\section{Comparison Against Numerical Simulations}
\label{Simulations}

We describe in appendix \ref{Numerical implementation} the numerical 
implementation of our model for $\cP(\rho_R)$. In particular, we explain how
to deal with the singularity $y_s$ of the generating function $\varphi(y)$
which appears on the negative real axis. The code to evaluate $\cP(\rho_R)$
can be downloaded from http://www-spht.cea.fr/pisp/valag/codepdf-en.html.

On the other hand, we investigate the robustness of our model in appendix
\ref{Robustness of the model}. That is, we study the dependence of our results
on the functional form (\ref{tauzeta}) we chose for $\tau(\zeta)$, which was
not fully determined by the quasi-linear and rare-void limits.

Then, we compare in this section the predictions of our model with the results
obtained from numerical simulations.

\subsection{Simulation properties}
\label{Simulation properties}

\begin{figure}
\protect\centerline{
\epsfysize = 3.truein
\epsfbox[277 149 588 470]
{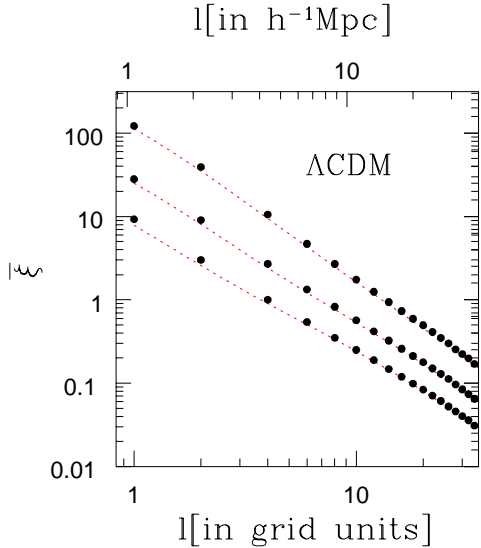} }
\caption{The variance computed from the simulations is compared with the 
analytical predictions (which follow Peacock \& Dodds 1996, see text) over 
a logarithmic scale. The solid dots are results from numerical simulations 
whereas lines are analytical predictions. Curves and dots from top down to 
bottom are for redshifts $z=0$, $z=1$ and $z=2$.}
\label{moment_var.eps}
\end{figure}

We have used the freely available intermediate scale $\Lambda$CDM N-body
simulation from 
Virgo Consortium\footnote{http://www.mpa-garching.mpg.de/Virgo/} to test our 
analytical predictions. The cosmological parameters are largely compatible 
with recent observations, in particular we have $\Om=0.3$, $\Ol=0.7$,
the Hubble constant is $H_0 = 70$ km/s/Mpc and $\sigma_8=0.9$. The box size 
is $L_{\rm box} = 239.5h^{-1}$Mpc while the number of particles is 
$N_{p} = 256^3$. We used three epochs at redshifts $z=0,1,2$ to compare our 
predictions with the simulations. We refer the reader to Jenkins et
al. (1998) and Thomas et al. (1998) for more details about the simulations.
The huge dynamic range studied by the VIRGO simulation provides an unique 
opportunity to probe the length scales from the quasi-linear regime up to the
highly non-linear regime.

To confront analytical results with numerical simulations we have used
a cubic $256^3$ grid. Then, we select cubic cells with volumes 
$l_{\rm grid}^3$, $8l_{\rm grid}^3$, $64l_{\rm grid}^3$ and 
$512l_{\rm grid}^3$. We compute the occupancy 
of each cell and the resulting count probability distribution. To increase 
the sampling random shifts were given to the 3D grid in orthogonal directions. 
This whole process was repeated several times to reach the number of cells 
which is required to probe low levels of probability. Note that an alternative
method which is equivalent to infinite sampling is developed in 
Szapudi (1998).  Next, we compare these numerical results with our analytical 
predictions. Note that our analytical pdfs apply to spherical cells (for 
which exact analytical results can be derived in the quasi-linear and 
rare-void limits). Therefore, we compare the pdf obtained from the N-body 
simulation for cubic cells of volume $l^3$ with our prediction for spherical 
cells of radius $R$ with the same volume: $l^3 = 4\pi R^3/3$. We shall see 
that we obtain a good agreement which shows that such a change in the shape 
of elementary cells has no strong influence. On the other hand, spurious 
effects such as the finite size of the simulations, shot noise and sampling 
errors need to be kept in mind while comparing analytical and simulation 
results.

Finite volume effects are due to the fluctuations of the underlying
density perturbations on scales larger than the simulation box. Typically
the tails of the count in cell (CIC) statistics are affected by the 
finite size of the catalogue. Such effects have been a very active area 
of research in recent past (Szapudi et al. 2000; Colombi et al. 2000) 
and much of the effort has been concentrated on trying to understand
how they can be ``corrected'' so that meaningful statistical indicators
can be constructed. Thus, it is known that the high-density far tail is
very sensitive to the presence (or absence) of rare clusters in the catalogue. 
This yields random fluctuations before the pdf shows an abrupt cutoff 
corresponding to the densest cell in the finite catalog. This can also give
a biased estimation for low order moments of the pdf such as the variance 
and the skewness. Of course, these problems worsen for smaller catalogs.
In our studies we have used of the order of $10^{10}$ cells of various sizes. 
This means that we can construct the probabilities $P_N$ down to $10^{-10}$.
We have avoided diluting the sample and we have used the full set of $256^3$
particles from the simulations in our construction of the pdf.

Discreetness effects are due to the sampling of the underlying continuous
density field with finite point sets. However, techniques to subtract the
Poisson shot noise from computed quantities are well known (see Munshi 
et al. 1999a  for expressions regarding the shot noise contributions 
to underlying cumulants of smoothed field from central moments and factorial 
moments of cell count statistics, see also Szapudi \& Szalay 1996). It is 
also possible to construct the analytical discrete pdf by convolving it 
with the Poisson sampling:
\beq
\cP_N = \int_0^{\infty} \d\rhoR  \; \cP(\rhoR) \;  \frac{(\rhoR\Nb)^N}{N!} \; 
e^{-\rhoR\Nb} , \;\;\; \Nb=\nb V ,
\label{pdf_poisson}
\eeq
where $\Nb$ is the mean number of particles in a cell (and $\nb$ is the mean
number density of particles in the simulation box). However, in most of our 
cases we found that this correction is not required as theoretical
predictions are already in good agreement with numerical data. On the other 
hand, since the lowest overdensity $\rhoR=N/\Nb$ that one can probe 
corresponds to $N=1$, which yields $\rhoR=1/\Nb$, larger cells which have
a larger $\Nb$ probe farther into the low-density tail of the pdf. For the 
computation of low order moments we relied on factorial moments which can 
directly be related to low order moments of the underlying continuous field: 
\beq
\sum_N \frac{N(N-1) \dots (N-p+1)}{\Nb^p} \cP_N 
= \int_0^{\infty} \d\rhoR  \; \cP(\rhoR) 
\; \rhoR^p .
\label{moment_poisson}
\eeq
For a complete discussion of generating functions, order by order expansions
of low order factorial moments and their link with the normalized 
cumulants parameters $S_N$ see Munshi et al. (1999a).

\begin{figure*}
\protect\centerline{
\epsfysize = 2.25truein
\epsfbox[25 301 588 603]
{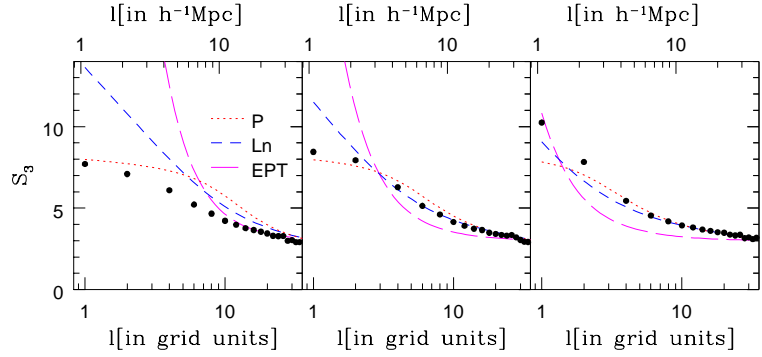} }
\caption{The skewness $S_3$ computed from the simulations is compared with
analytical predictions at redshifts $z=0,1,2$. Solid dots are results from 
numerical simulations whereas various lines are analytical predictions as 
labeled in the left panel. We display our interpolation eq.(\ref{S3}) (P), 
the EPT prediction from section~\ref{Extended Perturbation Theory} (EPT)
and the lognormal approximation eq.(\ref{S3ln}) (Ln).}
\label{moment_skew.eps}
\end{figure*}

\subsection{Variance and Skewness}
\label{Numerical Variance}

We first compare in Fig.~\ref{moment_var.eps} the variance $\xib(R,z)$ 
we obtain from the procedure described in section~\ref{Variance and skewness}
(following Peacock \& Dodds 1996) with the numerical simulations. We can see
that the agreement is quite good over all scales and redshifts we investigate
in this paper. Therefore, this simple procedure appears to be sufficient for
our purposes. Next, we display in Fig.~\ref{moment_skew.eps} the predictions
for the skewness $S_3(R,z)$ derived from our interpolation (\ref{S3}) between
the quasi-linear limit and HEPT (Scoccimarro \& Frieman 1999), from the EPT 
recalled in section~\ref{Extended Perturbation Theory} and from the lognormal
approximation (\ref{S3ln}), as compared with N-body simulations. We can check 
that all three models are very close on quasi-linear scales and match the
simulations. Note that for the lognormal model this is actually a coincidence
because its skewness only agrees with the exact quasi-linear limit for $n=-1$
(compare eq.(\ref{S3ln}) with eq.(\ref{S3QL})) and it happens that on the 
quasi-linear scales shown in Fig.~\ref{moment_skew.eps} we have $n\simeq -1$
(as is usual for CDM power-spectra at such redshifts). On the other hand, we
find that our interpolation and EPT prediction work rather well on 
non-linear scales at $z=1,2$ while the lognormal model shows a significant 
discrepancy. At $z=0$ our interpolation shows some deviation from the 
simulations but it fares much better than both EPT and the lognormal model. 
However, the comparison of the simulations at $z=0,1$ and $2$ suggests that 
on small scales finite resolution effects play a significant role (indeed, 
at fixed comoving scale one expects the skewness to grow with time, which 
is not the case in Fig.~\ref{moment_skew.eps} below a few grid lengths). 
Therefore, some of the discrepancy between numerical simulations and the 
models could be due to such numerical effects.

\subsection{Density pdf $\cP(\rhoR)$}
\label{Density pdf}

\begin{figure}
\protect\centerline{
\epsfysize = 3.5truein
\epsfbox[21 152 588 715]
{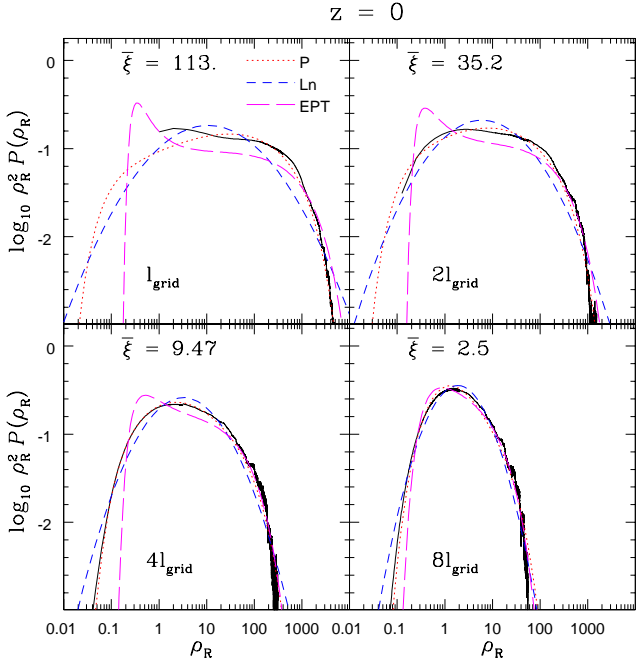} }
\caption{Analytical and numerical probability distribution functions are 
plotted for various smoothing scales $l$ (in grid units) as indicated 
in each panel. Lines of various styles represent analytical predictions from 
various theoretical models as labeled in the upper left panel. We show our
model (\ref{tauzeta}) (P), the EPT prediction (\ref{tauzetaEPT}) (EPT) and
the lognormal approximation (\ref{Pln}) (Ln). Dark solid lines are the results
from numerical simulations. We have plotted $\rhoR^2\cP(\rhoR)$ vs $\rhoR$ 
(on logarithmic scales) which shows clearly the evolution of the pdf into 
the non-linear regime.}
\label{pdf_z0_low.eps}
\end{figure}

\begin{figure}
\protect\centerline{
\epsfysize = 3.5truein
\epsfbox[21 152 588 715]
{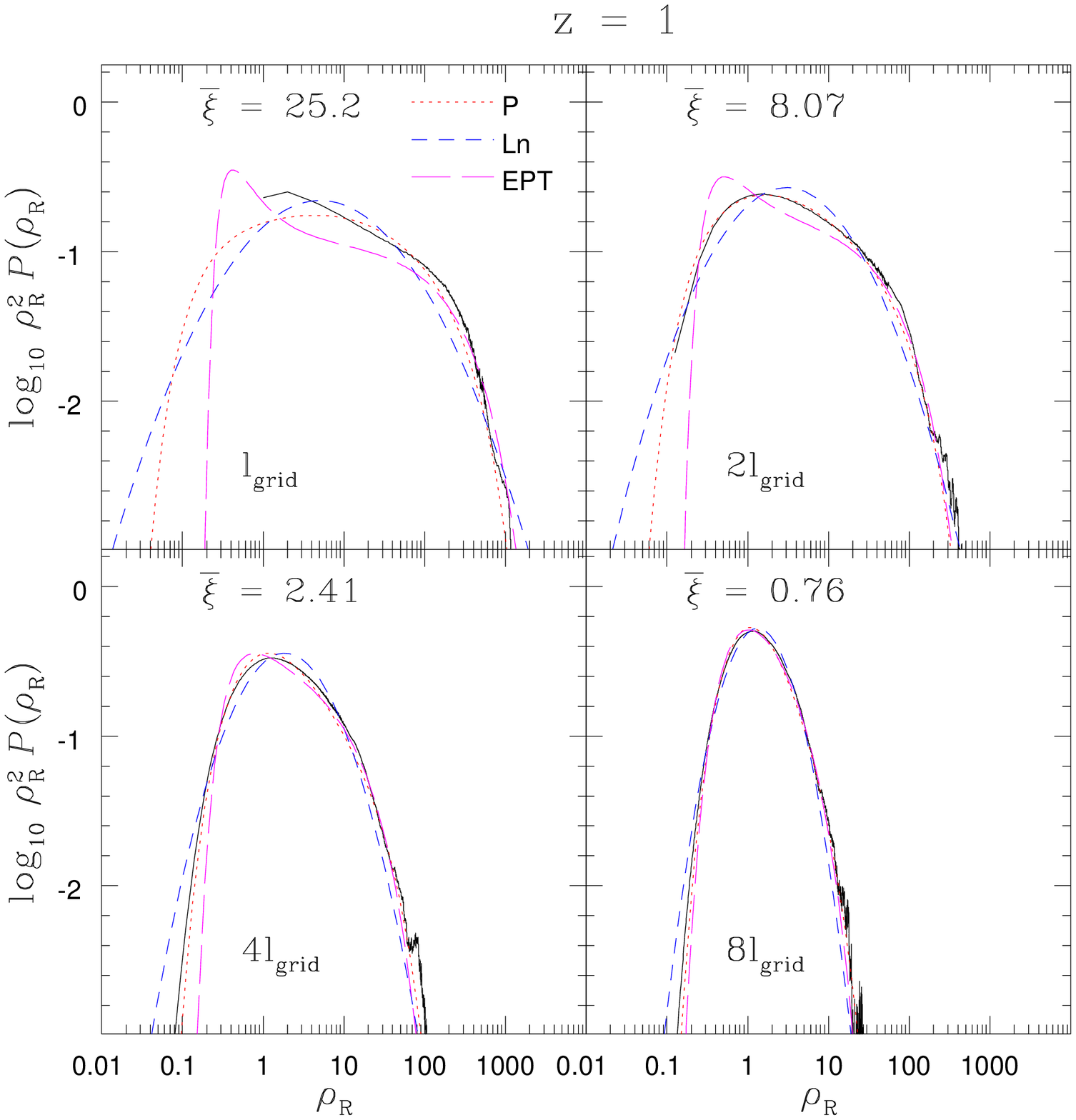} }
\caption{Same as previous figure but for z=1.}
\label{pdf_z1_low.eps}
\end{figure}

\begin{figure}
\protect\centerline{
\epsfysize = 3.5truein
\epsfbox[21 152 588 715]
{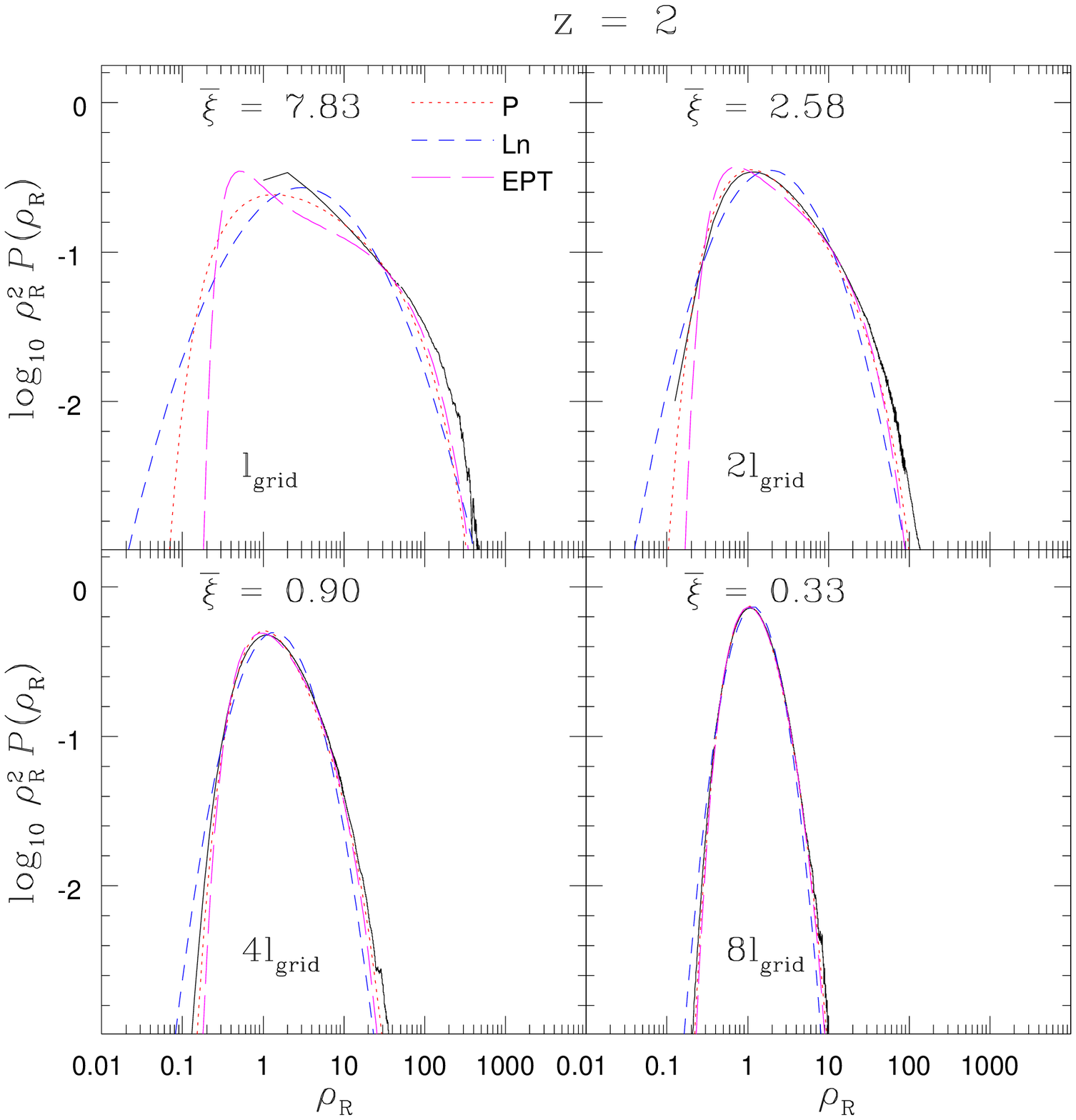} }
\caption{Same as previous figure but for z=2.}
\label{pdf_z2_low.eps}
\end{figure}

Finally, we compare in Figs.~\ref{pdf_z0_low.eps}-\ref{pdf_z2_low.eps} the
pdfs $\cP(\rhoR)$ obtained from the different theoretical models with the
results from N-body simulations. We display our model (\ref{tauzeta}), EPT
(\ref{tauzetaEPT}) and the lognormal model (\ref{Pln}) at redshifts $z=0,1$ and
$2$ and for four scales. We first plot $\rhoR^2 \cP(\rhoR)$ (in logarithmic
scales) which is the quantity of interest for most practical purposes. Indeed,
$\rhoR^2 \cP(\rhoR)$ is also the fraction of matter per logarithmic interval
of overdensity $\d\ln(\rhoR)$ at scale $R$. Moreover, it enables one to clearly
see the evolution of the pdf into the non-linear regime. Thus, we see that
while in the quasi-linear regime there is only one density scale:
the mean density of the universe (that is $\rhoR=1$ in our units) and the pdf
$\cP(\rhoR)$ tends to a simple Gaussian, as one goes deeper into the non-linear
regime two density scales gradually appear. A low-density scale $\rho_v \ll 1$
marks the low-density cutoff of the pdf, below which one enters the rare-voids
regime recalled in section~\ref{Rare voids}. As seen in Valageas (2002b),
this cutoff $\rho_v$ is given by:
\beq
\sigma \gg 1 : \;\;\; \rho_v= \sigma(R)^{-6/(1-n)} \ll 1 ,
\label{rhov}
\eeq
and the pdf $\cP(\rhoR)$ shows the modified exponential falloff:
\beq
\rhoR \ll \rho_v : \;\;\; \cP(\rhoR) \sim e^{-9 \rhoR^{-(1-n)/3}/(8\sigma^2)} .
\label{Pvoid}
\eeq
Note that in the quasi-linear regime (i.e. $\sigma \ll 1$) the low-density
falloff of $\cP(\rhoR)$ is still given by eq.(\ref{Pvoid}) but the density
scale $\rho_v$ goes to unity. This part of the pdf is governed by the last 
term in eq.(\ref{tauzeta}), that is by the behaviour (\ref{tauvoid}).
A second density scale $\rho_h$ which marks the high-density cutoff of the 
pdf $\cP(\rhoR)$ is simply given by the two-point correlation $\xib$:
\beq
\xib \gg 1: \;\;\; \rho_h = S_3 \; \xib = \frac{\lag\rhoR^3\rag_c}
{\lag\rhoR^2\rag_c} \gg 1 .
\label{rhoh}
\eeq
It corresponds to the typical overdensity within virialized halos at scale $R$
while higher densities are associated with rare massive objects. The reason
why we defined $\rho_h$ in eq.(\ref{rhoh}) from the second and third cumulants
of the density pdf is that they are the lowest order cumulants which are mostly
sensitive to the high-density part of the pdf. Thus, we can see from the 
figures that the mean $\lag\rhoR\rag=1$ probes the whole range 
$\rho_v<\rhoR<\rho_h$ and we can check that at small scales the cutoff $\rho_h$
can indeed be significantly higher than $\xib$ (compare 
Fig.~\ref{pdf_z0_low.eps} with Fig.~\ref{moment_var.eps}) following the 
growth of the skewness $S_3$.
On the other hand, higher-order cumulants probe increasingly large and rare
overdensities and exhibit higher uncertainties. As for the low-density cutoff
$\rho_v$, the high-density cutoff $\rho_h$ goes to unity in the quasi-linear
regime so that both density scales merge to the mean density of the universe.

We can see in Figs.~\ref{pdf_z0_low.eps}-\ref{pdf_z2_low.eps} that our simple
model (\ref{tauzeta}) agrees very well with the results from numerical 
simulations. Note moreover that our model has no free parameter beyond the
variance (obtained from Peacock \& Dodds 1996) and the skewness. The EPT
model recalled in section~\ref{Extended Perturbation Theory} is also fully
parameterized by $\xib$ and $S_3$, but we can see that it shows significant
discrepancies at small scales. This can be traced to its low-density behaviour
which does not obey the properties (\ref{tauvoid}) and (\ref{Pvoid}). Indeed,
it yields a much sharper low-density cutoff. In order to fulfill the 
normalization conditions (\ref{XiS3}) this implies a high peak at $\rho_v$
(to compensate for the smaller contributions from lower densities to the
normalization $\lag 1\rag=1$) and a lower plateau at intermediate densities
(to compensate for this high peak at $\rho_v$ into the normalization 
$\lag \rho_R \rag=1$). This failure to follow the evolution of the pdf into
the non-linear regime comes from the fact that the EPT model (\ref{tauzetaEPT})
attempts to parameterize the evolution of rare underdensities and overdensities
in the same fashion. That is, both the small $\zeta$ ($\zeta\ll 1$) and large
$\zeta$ ($\zeta \geq 1$) parts of the function $\tau(\zeta)$ are parameterized
by the same parameter $\neff$. In particular, if the stable-clustering Ansatz
were valid (that is $S_3$ goes to a constant with time in the non-linear 
regime at fixed physical scale) the function $\tau(\zeta)$ from EPT would 
become constant. 
However, it is clear that this would miss the physics at work at low 
densities. Indeed, even though high-densities are governed by virialization 
processes which might stabilize high-density peaks observed in physical units, 
rare voids are governed by different phenomena and keep expanding faster than 
the mean universe. This is captured by the last term in our model 
(\ref{tauzeta}) which shows the explicit time dependence $\xib/\sigma^2$.
This point is further discussed in section~3.6.3 in Valageas (2002b). Note 
that neither our model nor EPT assume that the stable-clustering Ansatz is 
valid but this discussion enables one to clearly see that rare underdensities 
and overdensities require distinct treatments.

The simple lognormal model shows a smooth parabolic-like shape in the scales 
used in Figs.~\ref{pdf_z0_low.eps}-\ref{pdf_z2_low.eps} (actually 
$\log[\rho_R \cP(\rho_R)]$ would give an exact parabola over $\log\rho_R$)
and it might appear to work
better than EPT. However, we can see that our model (\ref{tauzeta}) provides
significantly better results as compared with numerical simulations. Indeed,
the lognormal model yields low-density and high-density falloffs which are
clearly too shallow and it does not follow very well the flat plateau 
in-between $\rho_v$ and $\rho_h$ because of its parabolic shape (the curvature
is too high). The rather good success of our simple model (\ref{tauzeta}) is
due to its correct description of the low-density cutoff (as compared with EPT)
and the normalization constraints (\ref{XiS3}). Indeed, the rare-void limit
(\ref{tauvoid}) together with the normalization $\lag 1 \rag=1$ set the 
location and the height of the low-density cutoff $\cP(\rho_v)$ (as well as
lower densities). Next, the normalizations $\lag\rhoR^2\rag_c= \xib$ and
$\lag\rhoR^3\rag_c=S_3\xib^2$ set the location and the height of the 
high-density cutoff $\cP(\rho_h)$ (see eq.(\ref{rhoh})). Finally, the 
normalization $\lag\rhoR\rag=1$ provides a further constraint on the pdf
$\cP(\rhoR)$ over the whole range $\rho_v<\rhoR<\rho_h$. All these conditions
are sufficient to provide a tight constraint onto the pdf $\cP(\rhoR)$
at densities $\rhoR < \rho_h$. This property is obviously a direct consequence
of the weak dependency of the functions $\tau(\zeta)$ and $\varphi(y)$ onto 
additional free parameters discussed in appendix~\ref{Robustness of the model}
once all constraints (\ref{tauzetaQL})-(\ref{taunorm}) are taken into account.

It might be possible to have a significantly different pdf $\cP(\rhoR)$ while
satisfying the constraints (\ref{tauzetaQL})-(\ref{taunorm}) if we allow
large oscillations in the intermediate range $\rho_v<\rhoR<\rho_h$. However,
this would require a rather contrived model and
Figs.~\ref{pdf_z0_low.eps}-\ref{pdf_z2_low.eps} show that results from N-body
simulations do not exhibit such peculiar behaviour. On the contrary, although
the numerical simulations may not go sufficiently far into the non-linear
regime to draw definite conclusions, it seems that no intermediate density
scale shows up in the range $\rho_v<\rhoR<\rho_h$ and the pdf exhibits a
smooth power-law behaviour over this range. In this sense, the outcome of
gravitational clustering appears quite simple. Therefore, we can conclude
that our model is very robust over the whole range $\rhoR<\rho_h$. Moreover,
its prediction for $\cP(\rhoR)$ is the simplest one which can be made 
consistent with all known constraints. 

Any deviation from our result over 
this range (in fact over $\rho_v<\rhoR<\rho_h$ since lower densities are 
known from section~\ref{Rare voids}) would require additional parameters
which would for instance introduce new density scales. An alternative 
possibility would have been a bimodal distribution. For instance, in a fashion
similar to the behaviour shown by EPT in Fig.\ref{pdf_z0_low.eps}, one could
have imagined keeping the only two characteristic density scales $\rho_v$ and 
$\rho_h$ but having a specific power-law in the intermediate range which only 
connects smoothly to one extremity (for instance $\rho_h$) while the matching 
at the other hand (then $\rho_v$) involves a sharp transition. Another 
possibility would have been two distinct power-laws attached to each boundary 
$(\rho_v,\rho_h)$ which match in-between. It is not clear whether such 
behaviours could have been ruled out a priori. However, one may argue in this
direction as follows. If we assume that the stable clustering Ansatz is
valid, virialized objects are frozen in physical coordinates so that the
high-density part of the pdf remains constant. Then, the intermediate 
power-law range only grows towards underdensities as increasingly underdense
and extreme voids (as measured in the initial conditions) fill the whole volume
and see their fast expansion stopped as they join. In this process, the matter
at their boundaries forms filaments and virialized objects which become part
of the matter described by the power-law regime while the mass associated with
rare underdensities below $\rho_v$ keeps declining. Then, since at the
beginning of this process the two density scales $\rho_v$ and $\rho_h$ coincide
and the process repeats itself identically with time it is natural to expect
the build-up of a unique power-law regime which smoothly connects both 
end-points $(\rho_v,\rho_h)$ (while a double power-law could have been expected
if the matter in this range would originate both from $\rho_v$ and $\rho_h$,
the flux from one extremity increasing as one gets closer). In practice,
we do not expect such a stable clustering Ansatz to be valid, as moderate
overdensities experience mergings and tidal effects. Nevertheless, the 
scale-free nature of gravity and the approximate self-similarity of the 
process, which starts from a unique origin ($\rho_v=\rho_h$), could still
be the source of the simple behaviour seen in 
Figs.~\ref{pdf_z0_low.eps}-\ref{pdf_z2_low.eps}.

We must point out that this intermediate
density range actually corresponds to the part which is most difficult to
follow from a Lagrangian point of view. Indeed, it covers a wide range of
objects (about three orders of magnitude over the density in the upper panels
of Fig.~\ref{pdf_z0_low.eps}), from low-density filaments which surround large
voids up to typical virialized halos. The common property of these objects
is that they have undergone strong interactions with the neighbouring
density fluctuations and strong mergings. This makes it difficult for simple
Lagrangian mappings (which attempt to map the linear density contrast 
$\delta_L$ to the actual non-linear density contrast $\delta$, possibly with
some scatter) to model this density range. It is clear that our approach is
completely different from this point of view as we do not try to identify
non-linear density fluctuations from the linear density field on a one-to-one
basis. On the contrary, we directly focus on the pdf $\cP(\rhoR)$, which is
a statistical quantity, and we try to build a simple model which satisfies
all known constraints. As seen above, it appears that this is actually
sufficient to derive the pdf $\cP(\rhoR)$ over this density range up to
a good accuracy. As noticed above, the reason for this pleasant result is 
that the intricate merging process which ``shuffles'' the matter associated
with these typical density fluctuations does not bring about new scales and
it yields a simple power-law behaviour in-between the low and high density
cutoffs $\rho_v$ and $\rho_h$. Therefore, while the gravitational dynamics
of these individual objects is extremely complex (they may even lose their
identity through mergings or disruptions) the statistical outcome is very
simple. This suggests that the appropriate method to investigate this system
in a theoretical and rigorous manner from the equations of motion should
rely on a statistical analysis.

On the other hand, we must point out that the very high-density regime 
$\rhoR \gg \rho_h$ is not constrained by the conditions (\ref{XiS3})
which allowed us to obtain $\cP(\rhoR)$ for $\rhoR<\rho_h$ with a good
accuracy. As recalled in eq.(\ref{ys}), our 
model usually yields a singularity $y_s<0$ for the generating function 
$\varphi(y)$ (at least for $n<0$) which translates into a simple exponential 
cutoff for the high-density tail of $\cP(\rhoR)$, see eq.(\ref{expys}).
Nevertheless, we must note that this behaviour should not be taken at face 
value and the very high-density tail could exhibit a different shape.
As seen in Valageas (2002a) (section~3.6), this is actually the case in the 
quasi-linear limit where the singularity $y_s$ is somewhat spurious and one 
must follow a second branch of $\varphi(y)$ beyond $y_s$ which yields a 
high-density tail $\cP(\rhoR) \sim e^{-\rhoR^{(n+3)/3}/\sigma^2}$ which is
shallower than a simple exponential for $n<0$. As discussed in Valageas (2002b)
one can expect a similar behaviour at very high densities in the non-linear
regime. However, at the scales and redshifts shown in 
Figs.~\ref{pdf_z0_low.eps}-\ref{pdf_z2_low.eps} it appears that our model
works reasonably well up to the highest densities probed by the numerical
simulations and even higher densities are probably too rare to be of practical
interest. On the other hand, in the quasi-linear regime our model goes to
the exact quasi-linear limit (like EPT). As seen in Valageas (2002a), in this
regime too the far tail where the deviation from the simple exponential cutoff
discussed above could be seen corresponds to very rare events which are 
usually irrelevant.

\begin{figure}
\protect\centerline{
\epsfysize = 3.5truein
\epsfbox[21 152 588 715]
{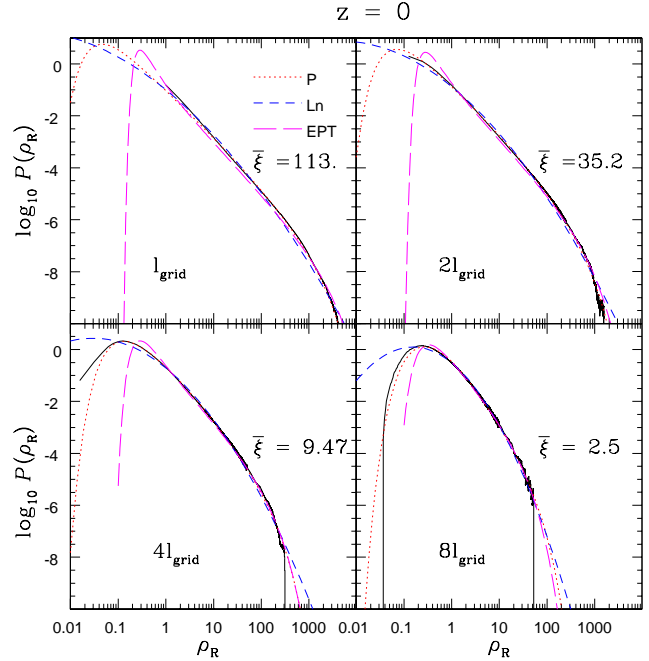} }
\caption{Analytical and numerical probability distribution functions
are plotted for various smoothing scales $l$ (in grid units) as 
indicated in each panel, over logarithmic scales. Lines of various styles
represent analytical predictions from various theoretical models, while
the numerical data is shown by dark solid lines, as in 
Fig.~\ref{pdf_z0_low.eps}.}
\label{pdf_z0.eps}
\end{figure}

\begin{figure}
\protect\centerline{
\epsfysize = 3.5truein
\epsfbox[21 152 588 715]
{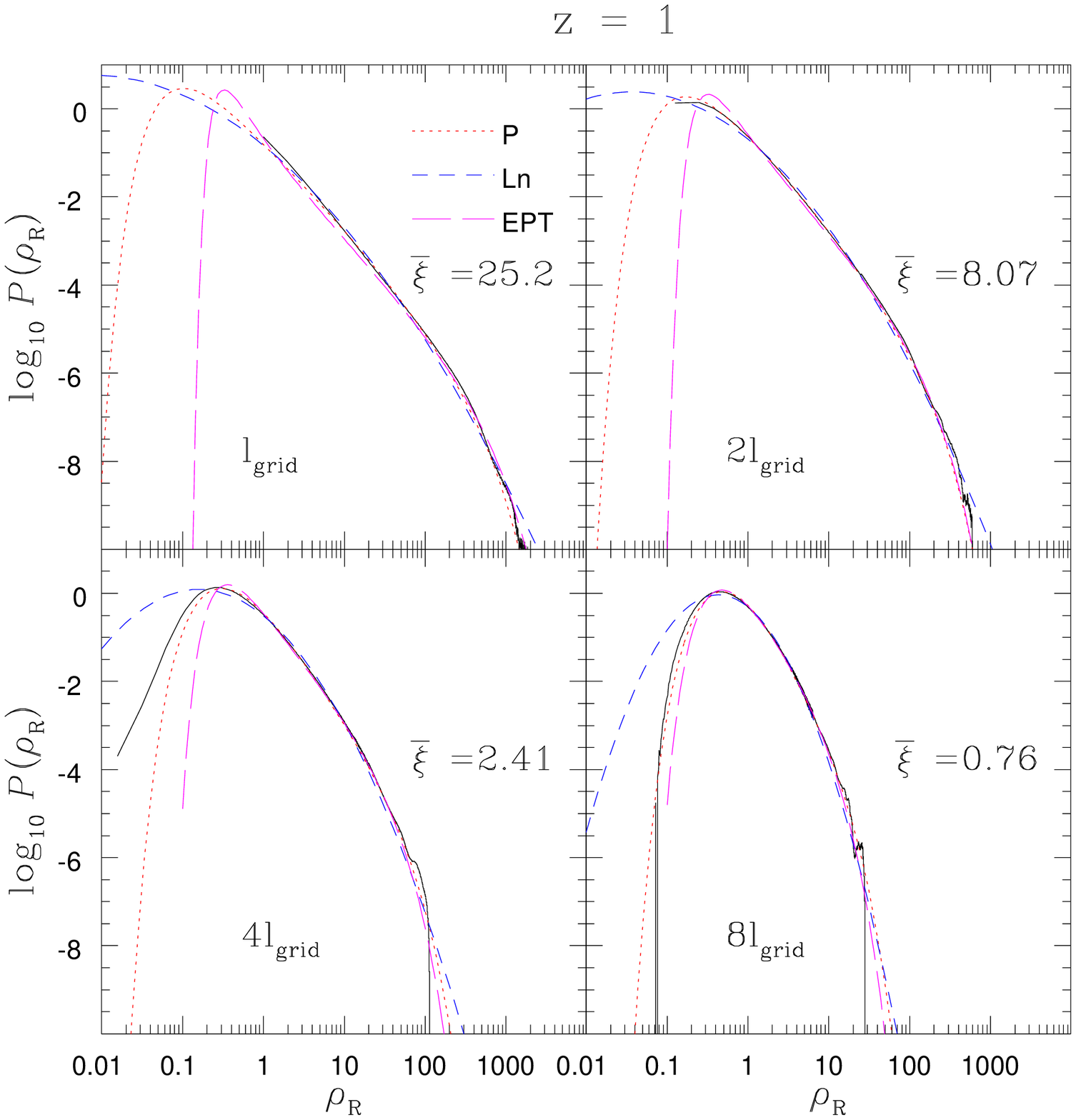} }
\caption{Same as previous figure but for z=1.}
\label{pdf_z1.eps}
\end{figure}

\begin{figure}
\protect\centerline{
\epsfysize = 3.5truein
\epsfbox[21 152 588 715]
{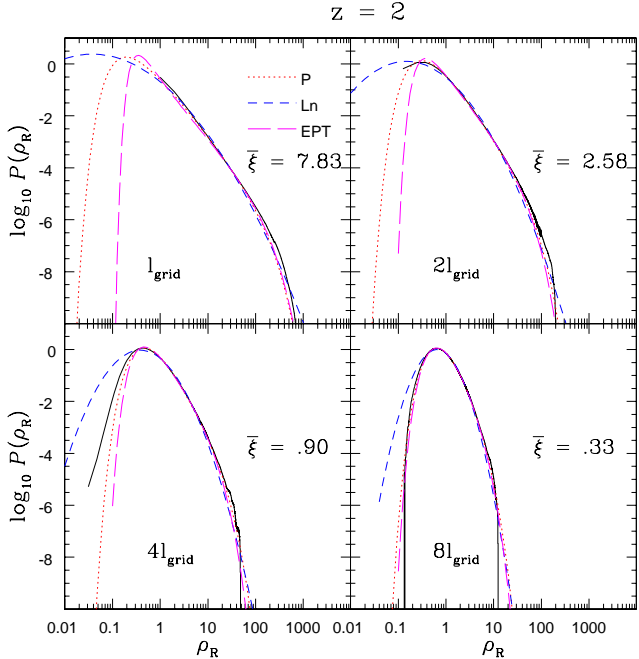} }
\caption{Same as previous figure but for z=2.}
\label{pdf_z2.eps}
\end{figure}

Finally, we show for completeness in Figs.~\ref{pdf_z0.eps}-\ref{pdf_z2.eps}
the pdf $\cP(\rhoR)$ from the various theoretical models and the N-body
simulations. This emphasizes the low-density cutoff $\rho_v$ but the 
high-density cutoff $\rho_h$ is blurred by the huge vertical scale and cannot
be easily distinguished from the intermediate power-law part. Of course,
as in Figs.~\ref{pdf_z0_low.eps}-\ref{pdf_z2_low.eps} we can check that our 
model (\ref{tauzeta}) works best, while EPT yields too sharp a low-density
cutoff at a larger density and the lognormal model gives too shallow a cutoff
at a smaller density.

\section{Conclusion}
\label{Conclusion}

Thus, in this paper we have presented a new model to describe the evolution
of the density probability distribution function $\cP(\rhoR)$. This allows
us to follow the dynamics of gravitational clustering on cosmological scales
from the linear regime up to the highly non-linear regime. Taking advantage of
the known rigorous results which have been obtained in the quasi-linear limit
and the rare underdense limit, we have built the simplest model which satisfies
both these constraints as well as normalization conditions. Our model is fully
parameterized by the non-linear variance and the skewness. Using standard
estimates for these two quantities we have shown that our predictions match 
N-body simulations over the scales and redshifts of interest. Moreover, we
have found that our model works significantly better than previous 
approximations (EPT and the lognormal).

We have explained that this success is due to two properties: i) the correct
handling by our model of the rare void limit and ii) the simple statistical
outcome of gravitational clustering over the range spanned by typical 
fluctuations. Indeed, following previous works (e.g., Balian \& Schaeffer 1989,
Colombi et al. 1997, Valageas 1998,1999) we have checked that in the non-linear
regime two distinct density scales appear (which merge to the mean density of 
the universe $\rhob$ in the linear regime): a low density cutoff 
$\rho_v \ll \rhob$ and a high-density cutoff $\rho_h \gg \rhob$. The 
low density cutoff $\rho_v$ marks the transition to rare voids which keep
expanding faster than the Hubble flow while the high density cutoff $\rho_h$
marks the transition to rare overdensities. The intermediate range 
$\rho_v<\rhoR<\rho_h$ corresponds to typical density fluctuations which have 
undergone mergings and tidal disruptions and spans a large variety of objects:
from filaments surrounding large voids to moderate virialized halos. These
structures cannot be followed by Lagrangian approaches since they have no well
defined identities (because of mergings and disruptions) but it appears that
the statistical outcome of this intricate gravitational dynamics is quite 
simple: the pdf obtained from numerical simulations exhibits a simple 
power-law behaviour over the whole range $\rho_v<\rhoR<\rho_h$ (i.e. no 
additional density scale appears). Then, this simple behaviour together with
the rare-void limit and the normalization conditions provide a tight 
constraint on the pdf $\cP(\rhoR)$ over $\rhoR<\rho_h$ which can thus be
predicted with a good accuracy. This also explains why it is possible to
build a satisfactory model using only $\xib$ and $S_3$. As described in
section~\ref{Density pdf}, these two quantities provide the location $\rho_h$
(see eq.(\ref{rhoh})) and the height $\cP(\rho_h)$ of the high-density
cutoff. Then, a simple power-law matching to the low-density cutoff $\rho_v$
(whose location and height are explicitly known as a function of $\sigma^2$)
allows a complete description at all densities below (and also of the order
of) $\rho_h$.

On the other hand, we have noticed that there are no rigorous results for the
very high-density limit $\rhoR\gg\rho_h$ so that our model should be viewed
with some caution in this regime. In particular, although our simple model
usually yields an exponential tail at high densities we explained that one
could expect a more general shape (the exponential of some power-law) as
discussed in Valageas (2002b). If this high-density falloff were sharper than
exponential it could be handled within our framework in a straightforward way
by modifying the large-$\zeta$ part of the function $\tau(\zeta)$ we used
in eq.(\ref{tauzeta}) and making sure it yields no singularity. By contrast,
if this high-density cutoff were shallower than exponential it would require
a more important modification (see also Valageas 2002a for a similar case).
A possible way to do so would be to consider the pdf and the cumulant 
generating function associated with the logarithm of the density $\ln\rhoR$
rather than the density $\rhoR$ itself. Here we may note that going to
higher orders over $S_p$ (i.e. adding further constraints to (\ref{XiS3})
by including higher-order moments) does not appear to us to be the most 
promising
way to improve this model. Indeed, it would add some further parameters
which are not accurately known without ensuring that the large density limit
is correct. In our view, one would rather like to derive the large-density
behaviour of the pdf $\cP(\rho_R)$, or of the generating function 
$\tau(\zeta)$, and build a simple model for $\tau(\zeta)$ which recovers this
asymptotic behaviour (as we did for the underdense limit). Unfortunately,
this behaviour has not been derived yet (although one might try a version
of the Press-Schechter (1974) prescription coupled to the stable-clustering 
Ansatz or some halo model). As discussed above, in such a case it could be 
more convenient to work with $\ln\rhoR$.

However, we shall not explore this 
point further in this paper because we have found that our model works
quite well over the scales and redshifts of interest. Moreover, the very
high-density tail $\rhoR\gg\rho_h$ where such deviations might appear
is beyond the reach of these simulations and it corresponds to very rare
overdensities which should be irrelevant for most practical purposes. In
particular, the robustness of our model discussed above for $\rhoR<\rho_h$
ensures that we obtain good predictions for density fluctuations which occupy
most of the volume and contain most of the matter of the universe. Besides,
we can note that the constraint provided by the skewness actually means that 
we correctly describe the near high-density tail.

Therefore, we expect that this simple model should be useful for cosmological
studies which require a realistic estimate of the probability distribution
$\cP(\rhoR)$, from rare voids up to rare overdensities and from the linear
regime up to the highly non-linear regime. In fact, this model is probably the
simplest one which can be built consistently with all known results. It also
suggests that a theoretical study of gravitational clustering on non-linear
cosmological scales should rely on a statistical analysis rather than 
a detailed Lagrangian approach, as the properties of the system seem to be
much simpler within a statistical perspective. However, such an analysis
still remains to be done.

Finally, although we have focused on the one-point pdf in this paper,
it is possible to extend our approach to analyze the bias associated
with overdense cells and the detailed description of their multivariate
distribution. A detailed analysis in line with Munshi et al. (1999b) and 
Bernardeau \& Schaeffer (1999) will be presented elsewhere.

\section*{acknowledgments}

DM was supported by PPARC of grant RG28936. We would like to thank Tom Theuns 
for helping us with the analysis of VIRGO simulation data.
It is a pleasure for DM to acknowledge many fruitful
discussions with members of Cambridge Leverhulme Quantitative
Cosmology Group. Simulations analysed here were carried out at 
Edinburgh Parallel Computing Centre as part of the Virgo 
Supercomputing Consortium.

\appendix

\section{Numerical implementation}
\label{Numerical implementation}

\begin{figure}
\begin{center}
\psfig{figure=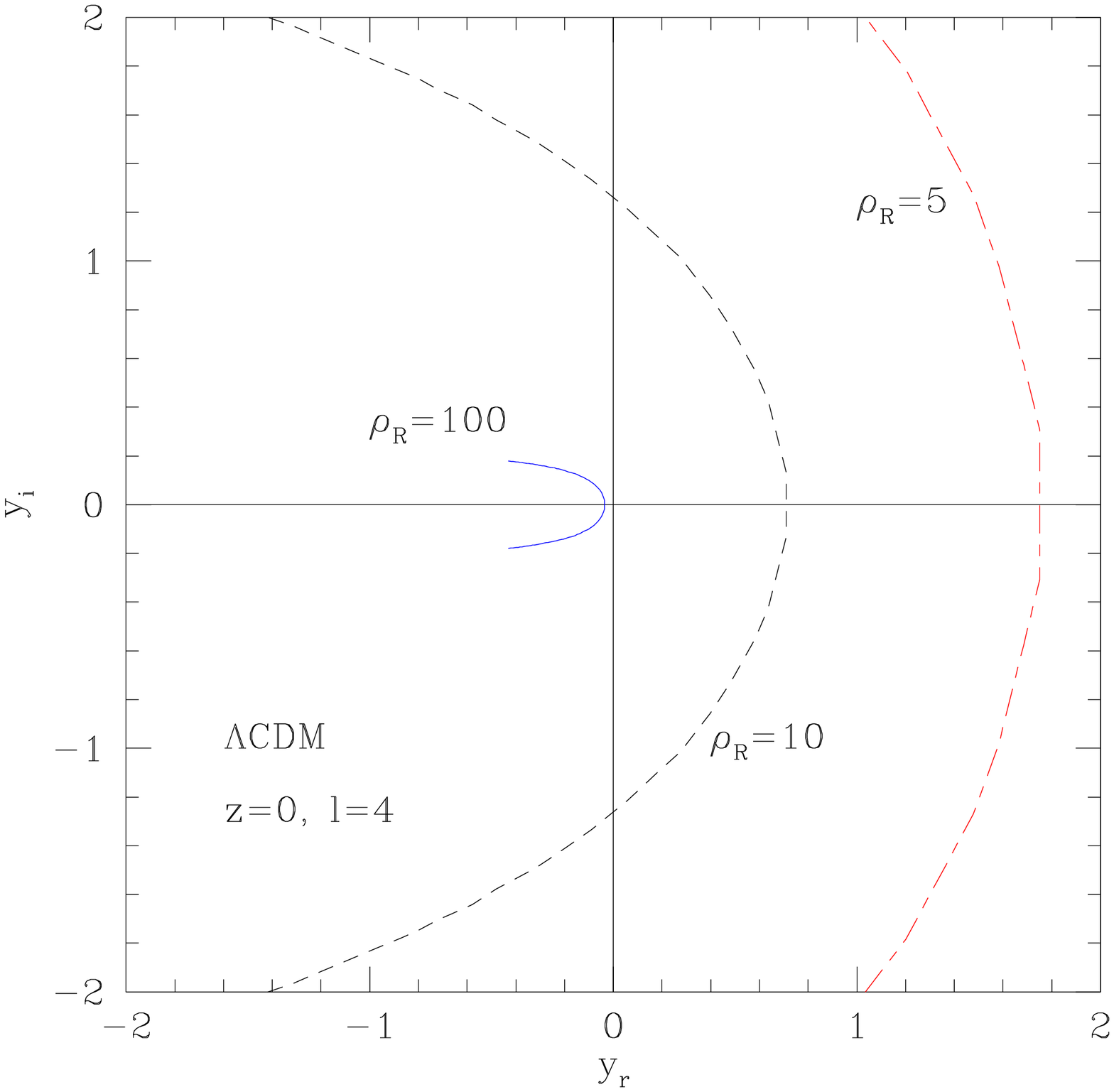,width=7cm,height=7cm}
\end{center}
\caption{The integration path over the complex $y-$plane used to compute the
pdf $\cP(\rhoR)$ from eq.(\ref{Pphiln}). We show the paths obtained at $z=0$
and at the scale $l=4 l_{\rm grid}$ for $\rho_R=100, 10$ and $5$ (this 
corresponds to the lower left panels in 
Figs.~\ref{pdf_z0_low.eps},\ref{pdf_z0.eps}).}
\label{Figypath}
\end{figure}

We have described in section \ref{Models} two models which yield the pdf
$\cP(\rhoR)$ through its cumulant generating function $\varphi(y)$, the latter
being determined through the implicit system (\ref{phizeta})-(\ref{ytau}). We
explain in more details here how to compute numerically $\cP(\rhoR)$ within
this framework. As is well-known, in the context of large scale structures
the functions $\tau(\zeta)$ which appear in the implicit system 
(\ref{phizeta})-(\ref{ytau}) usually yield a singularity $y_s$ on the negative 
real axis for $\varphi(y)$ because the function $\tau(y)$ is bivaluate (e.g., 
Bernardeau \& Schaeffer 1992, Bernardeau 1992, Valageas 2002a). This means that
the parametric representation (\ref{phizeta})-(\ref{ytau}) for $\varphi(y)$
yields two branches (obtained for $\tau>\tau_s$ and $\tau<\tau_s$) which join 
at $y_s$, as seen for instance in Figs.~3-5 in Valageas (2002a) (see also
Figs.~\ref{Figyzeta}-\ref{Figphiy} below). From the definition 
(\ref{phi})-(\ref{phiP}) of $\varphi(y)$, the branch
of interest is the one which runs through $(y=0,\varphi=0)$ and 
$(\zeta=1,\tau=0)$, where we have the Taylor expansion 
$\varphi(y)=y-y^2/2+..$ (as we noticed below eq.(\ref{tauhzetah}) the linear 
term $y$ is usually removed in studies which focus on the quasi-linear regime 
by working with $\deltaR$ rather than $\rhoR$). As explained in 
Valageas (2002a), the non-perturbative steepest-descent approach shows that 
in the quasi-linear limit the second branch of $\varphi(y)$ is actually 
relevant as it governs the high-density tail of the pdf $\cP(\rhoR)$. However,
for our present purposes it is irrelevant because our perspective is quite
different. In this paper we do not compute $\cP(\rhoR)$ from the equations of
motion, we merely build a phenomenological model for the pdf. Hence we actually
define $\varphi(y)$ from the implicit system (\ref{phizeta})-(\ref{ytau})
so that the singularity $y_s$ is no longer an artifact but an element of the
definition of $\varphi(y)$. As described in Colombi et al. (1997), one obtains
from eqs.(\ref{phizeta})-(\ref{ytau}) the behaviour of $\varphi(y)$ near its
singularity as:
\beq
y \geq y_s : \; \varphi(y) = \varphi_s + r_s (y-y_s) + a_s (y-y_s)^{3/2} 
+ .. ,
\label{ys}
\eeq
which implies through the inverse Laplace transform (\ref{Pphi}) the 
high-density behaviour:
\beq
\rhoR \gg 1+\xib : \;\; \cP(\rhoR) \propto \rhoR^{-5/2} \; 
e^{-|y_s|\rhoR/\xib} .
\label{expys}
\eeq
In practice, we must ensure that the numerical computation of the inverse 
Laplace transform (\ref{Pphi}) does not cross the branch cut $y<y_s$. 
Moreover, as explained in Colombi et al. (1997), it is important to choose
the integration path in the complex $y$-plane such that the argument of the
exponential in eq.(\ref{Pphi}) is real, which avoids oscillations and ensures
a fast convergence of the integral. To do so, one starts on the real axis at
$y=y_c$ where $y_c$ is the saddle-point of the exponent (the path is symmetric
with respect to the real axis). However, as recalled above we must not cross
the real axis below $y_s$, hence for $y_c < y_s$ one usually starts the
integration from $y_s$ (e.g., Colombi et al. 1997, Munshi et al. 2004).
However, this procedure is not fully satisfactory, especially in the highly
non-linear regime. We found that it is more efficient to first
perform two integrations by parts in eq.(\ref{Pphi}) which yields:
\beq
\cP(\rhoR) = \frac{1}{\rhoR^2} \inta \frac{\d y}{2\pi i} \; 
e^{[\rhoR y - \varphi + \xib \ln(\varphi'^2/\xib-\varphi'') ] /\xib} .
\label{Pphiln}
\eeq
Then, the saddle-point is given by:
\beq
\rhoR = \varphi' - \xib \; \frac{\varphi''' - 2 \varphi' \varphi''/\xib}
{\varphi'' - \varphi'^2/\xib} \;\;\;\; \mbox{at} \;\;\;\; y_c .
\label{yc}
\eeq
Now the saddle-point $y_c$ ``feels'' the existence of the singularity $y_s$
and always obeys $y_c>y_s$. Indeed, from eq.(\ref{ys}) one can easily see that
we have the asymptotic behaviour:
\beq
\rhoR \rightarrow \infty : \;\; y_c-y_s \sim \frac{\xib}{\rhoR} ,
\label{ycys}
\eeq
which automatically gives the high-density behaviour (\ref{expys}) from
eq.(\ref{Pphiln}). Note that a single integration by parts would also
give $y_c>y_s$ but it would yield the spurious scaling $(y_c-y_s) \sim 
(\xib/\rhoR)^2$ which would naively imply a power-law prefactor $\rhoR^{-3}$
instead of $\rhoR^{-5/2}$ in eq.(\ref{expys}). Of course, in principle any
procedure (with none or several integrations by parts) must give the same
results but those which are not well-suited to the singular behaviour 
(\ref{ys}) imply a higher computational cost for a given accuracy. On the
other hand, the underdense regime ($\rhoR \rightarrow 0$ and $y_c \rightarrow 
+\infty$) shows no particular problem. As seen in Valageas (2002b) the
exact asymptotic behaviour (\ref{tauhvoid}), which is verified by our model,
leads to:
\beq
y \rightarrow +\infty : \;\; \varphi(y) \propto y^{(1-n)/(4-n)} ,
\label{phiyinf}
\eeq
and:
\beq
\rhoR \rightarrow 0 : \;\; \cP(\rhoR) \propto \rhoR^{(n-13)/6} \; 
e^{-9 \rhoR^{-(1-n)/3} /(8\sigma^2)} .
\label{Pvoid}
\eeq
The numerical factor $9/8$ in eq.(\ref{Pvoid}) was changed to $(27/20)^2/2$
in Valageas (2002b) where we considered a critical-density universe. As
noticed in section \ref{Models}, this dependence on cosmology can actually be 
neglected.
Moreover, one easily see from eq.(\ref{Pphi}) that the asymptotic behaviour
(\ref{phiyinf}) implies $\cP(\rhoR) = 0$ for $\rhoR<0$ (by pushing the 
integration path over $y$ to $\R(y) \rightarrow +\infty$).

We show in Fig.\ref{Figypath} the integration paths over the complex $y-$plane
used to compute the pdf $\cP(\rhoR)$ from eq.(\ref{Pphiln}), at $z=0$
and at the scale $l=4 l_{\rm grid}$, for $\rho_R=100, 10$ and $5$, which 
corresponds to the lower left panels in 
Figs.~\ref{pdf_z0_low.eps},\ref{pdf_z0.eps}. For lower densities the 
integration path crosses the real axis at a larger $y_c$ (with 
$y_c \rightarrow +\infty$ for $\rhoR \rightarrow 0$) while for high densities
it gets closer to the singularity $y_s<0$ (with $y_c \rightarrow y_s^+$ for
$\rhoR \rightarrow \infty$, see eq.(\ref{ycys})) and it bends more closely 
along the branch cut (i.e. the negative real axis with $y<y_s$). Moreover,
for the same relative accuracy, for a higher density one only needs to 
integrate over a shorter length on $y$ but with smaller steps.

\section{Robustness of the model}
\label{Robustness of the model}

\begin{figure}
\begin{center}
\psfig{figure=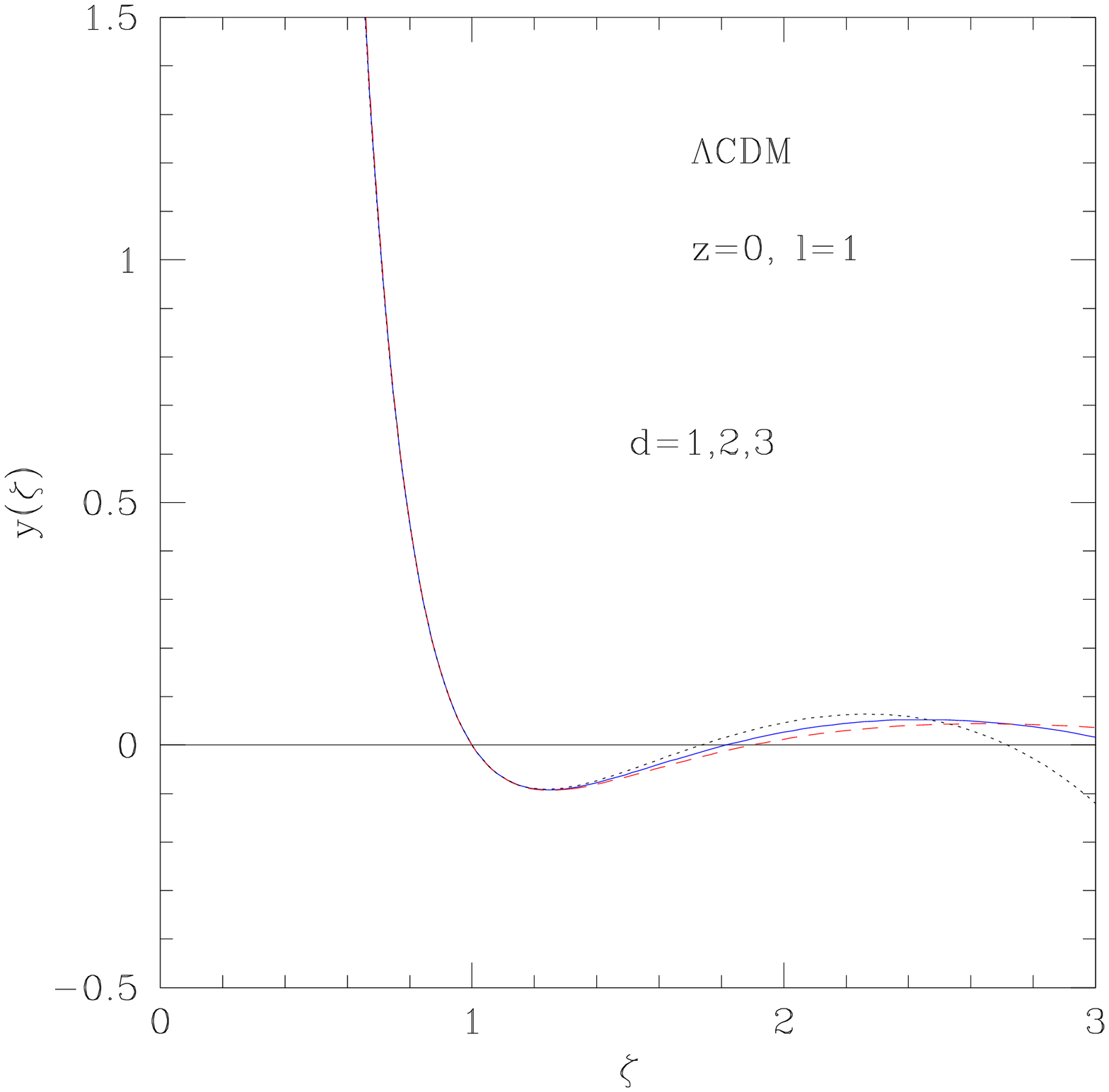,width=7cm,height=7cm}
\end{center}
\caption{The function $y(\zeta)$ obtained at $z=0$ and at the scale 
$l=l_{\rm grid}$ (this corresponds to the upper left panels in 
Figs.~\ref{pdf_z0_low.eps},\ref{pdf_z0.eps}). We show the results derived 
with $d=1$ (dashed line), $d=2$ (solid line) and $d=3$ (dotted line).}
\label{Figyzeta}
\end{figure}

\begin{figure}
\begin{center}
\psfig{figure=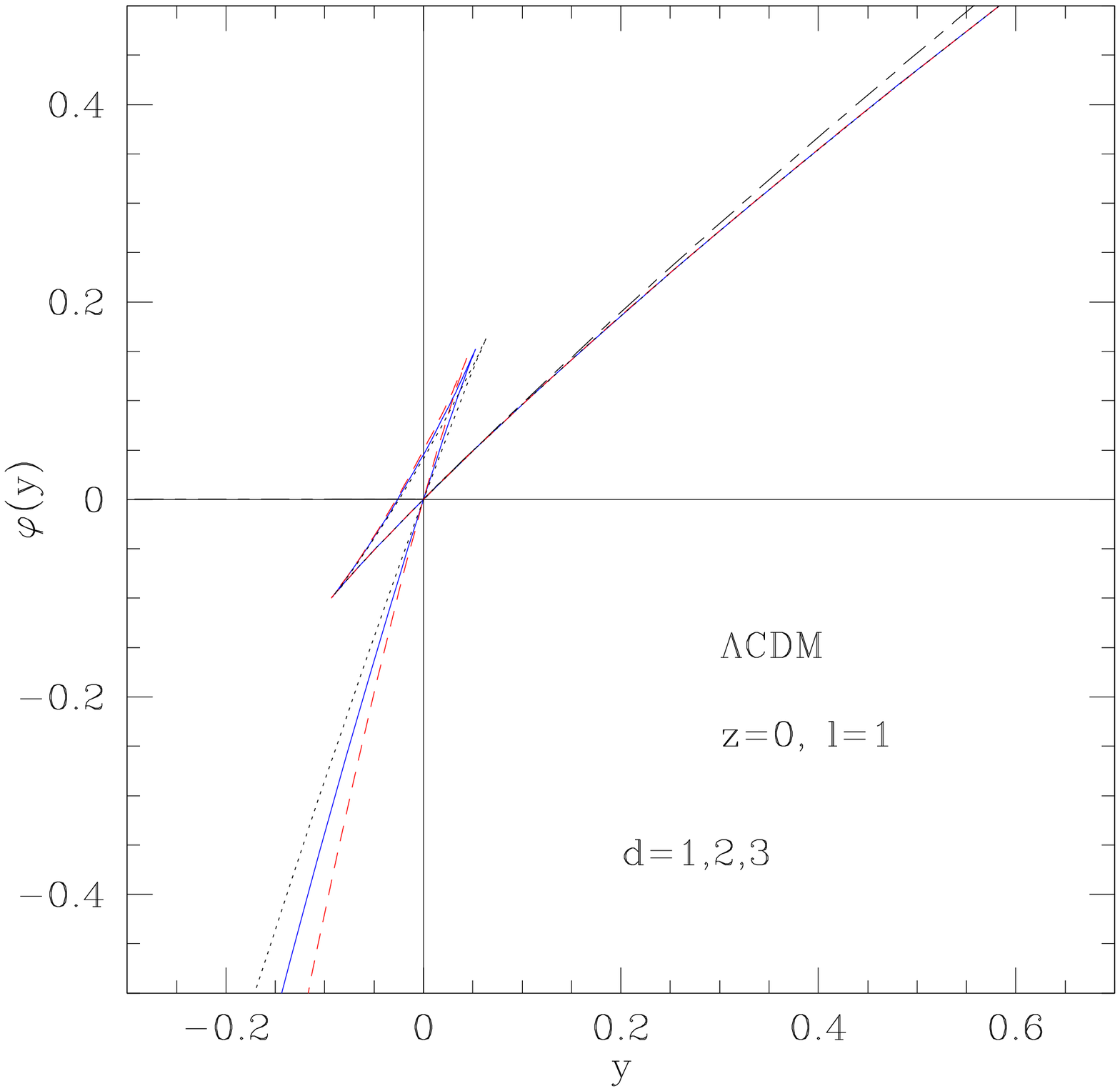,width=7cm,height=7cm}
\end{center}
\caption{The cumulant generating function $\varphi(y)$ obtained at $z=0$ and 
at the scale $l=l_{\rm grid}$ (this corresponds to the upper left panels in 
Figs.~\ref{pdf_z0_low.eps},\ref{pdf_z0.eps}). We show the results derived 
with $d=1$ (dashed line), $d=2$ (solid line) and $d=3$ (dotted line). We also
plot $\varphi(y)$ for the lognormal model (\ref{Pln}) (dot-dashed curve
for $y \geq 0$).}
\label{Figphiy}
\end{figure}

In the simple expression (\ref{tauzeta}) which defines our model, the last
two terms are set by the quasi-linear limit and the low-density regime. 
However, the first two terms $a+b\zeta^2$ are somewhat arbitrary. Indeed, 
although their normalization is given by the constraints (\ref{taunorm}) we
could have used other powers or functions. Thus, we could replace for instance
$a+b\zeta^2$ by $a+b\zeta^d$ with any $d$ larger than $(n-1)/6$ (so that the
low-density limit $\zeta\rightarrow 0$ remains correct). However, we have 
checked through numerical computations that the dependence of our results
on this free parameter is actually negligible: the different curves we obtain
for the pdf $\cP(\rhoR)$ are almost indistinguishable in all cases described in
section~\ref{Simulations}, where we compare our results with N-body
simulations. The reason for this independence on the parameter $d$ can be
seen in Fig.~\ref{Figyzeta}, where we display the function $y(\zeta)$ obtained
from eqs.(\ref{ytau}),(\ref{tauzeta}). We consider the scale $l=l_{\rm grid}$
(with $l_{\rm grid}$ being the grid unit of the simulations) and the redshift
$z=0$, which is the most non-linear case 
(upper left panels in Figs.~\ref{pdf_z0_low.eps},\ref{pdf_z0.eps}). 
We show in Fig.~\ref{Figyzeta} 
the functions $y(\zeta)$ obtained for a parameter $d=1,2$ and $3$, the exponent
$d=2$ being our fiducial model as written in eq.(\ref{tauzeta}). Then, we see
that all three curves $y(\zeta)$ are very close for $\zeta\la 1.5$. Indeed, 
the low-density part $\zeta<1$ goes to the rare-void limit described in 
section~\ref{Rare voids} and eq.(\ref{tauvoid}) while the behaviour near
$\zeta=1$ is constrained by (\ref{taunorm}). Next, we note that at 
$\zeta_s \simeq 1.3$ the function $y(\zeta)$ reaches a minimum. This 
translates into a singularity for $\varphi(y)$ since this means that $\zeta(y)$
and $\varphi(y)$ become bivaluate. More precisely, near this minimum we have
$y-y_s \sim (\zeta-\zeta_s)^2$ which yields 
$\zeta-\zeta_s \sim \pm |y-y_s|^{1/2}$
and the power $3/2$ in eq.(\ref{ys}) for $\varphi(y)$. Then, as discussed
in section~\ref{Numerical implementation}, this gives several branches for
$\varphi(y)$ and we must only keep the one which connects to the underdense
regime, which corresponds to the part of $y(\zeta)$ with $\zeta<\zeta_s$.
Since this singularity $\zeta_s$ is close to $1$ it is strongly constrained by
eqs.(\ref{taunorm}) and the dependence on the parameter $d$ of the function
$\tau(\zeta)$ (or $y(\zeta)$) is very weak over this range $\zeta<\zeta_s$.

This behaviour can also be seen from Fig.~\ref{Figphiy} where we display the
cumulant generating function $\varphi(y)$ for the same cases as in 
Fig.~\ref{Figyzeta}. The relevant branch of $\varphi(y)$ which defines the
pdf $\cP(\rhoR)$ is the one which runs through the origin $y=0$ and extends
up to $y \rightarrow +\infty$ (extreme underdensities). Again, we see that
the three curves obtained for $d=1,2$ and $3$ are indistinguishable over this
branch. By contrast, one can see some deviations appear as one follows the 
other branches down to $y\rightarrow-\infty$ (which corresponds here to $\zeta
\rightarrow +\infty$) but they have no signification for the pdf $\cP(\rhoR)$.

For completeness, we also display in Fig.~\ref{Figphiy} the generating 
function $\varphi(y)$ obtained for the lognormal model (\ref{Pln}). It is
restricted to $y \geq 0$ since the integral (\ref{phiP}) diverges for
$y<0$ because the high-density tail of the lognormal decreases more
slowly than a simple exponential. In other words, we now have $y_s=0$.
On the other hand, at very large $y$ the lognormal approximation yields
$\varphi(y) \sim \ln^2 y$ contrary to the power-law (\ref{phiyinf}).
Thus, one can already see on $\varphi(y)$ the differences between various 
models for $\cP(\rho_R)$.

\end{document}